\documentclass[a4paper,USenglish,cleveref, autoref]{lipics-v2021}
\pdfoutput=1

\usepackage[utf8]{inputenc}
\usepackage[T1]{fontenc}
\usepackage{fontaxes}
\usepackage{microtype}
\usepackage{mathrsfs}
\usepackage{mathtools}
\usepackage[defaultlines=3,all]{nowidow}
\usepackage{relsize}
\usepackage{graphicx}

\newcommand{\Cc}{\mathscr{C}}
\newcommand{\Dd}{\mathscr{D}}

\newcommand{\N}{\mathbb{N}}

\renewcommand{\phi}{\varphi}

\DeclareMathOperator{\sd}{\text{SC-depth}}
\renewcommand{\epsilon}{\varepsilon}

\newtheorem{problem}{Problem}
\newtheorem{fact}{Observation}

\DeclareMathOperator{\comp}{\oplus}

\DeclareMathOperator{\Frank}{{\rm Frank}}
\DeclareMathOperator{\coloring}{\mathfrak{c}}
\makeatletter

\newcounter{tmpthm}
\newcounter{thmmain}
\makeatother

\crefname{conjecture}{Conjecture}{Conjectures}
\crefname{theorem}{Theorem}{Theorems}

\title{Decomposition horizons and a characterization of stable hereditary classes of graphs} 

\author{Samuel Braunfeld}{Computer Science Institute of Charles University (IUUK), Praha, Czech Republic}{sbraunfeld@iuuk.mff.cuni.cz}{https://orcid.org/0000-0003-3531-9970}{}

\author{Jaroslav Ne\v set\v ril}{Computer Science Institute of Charles University (IUUK), Praha, Czech Republic}{nesetril@iuuk.mff.cuni.cz}{https://orcid.org/0000-0002-5133-5586}{}

\author{Patrice {Ossona de Mendez}}{Centre d'Analyse et de Math\'ematiques Sociales (CNRS, UMR 8557),
	Paris, France and\\
	Computer Science Institute of Charles University,
	Praha, Czech Republic}{pom@ehess.fr}{https://orcid.org/0000-0003-0724-3729
}{}

\author{Sebastian Siebertz}{University of Bremen, Bremen, Germany}{siebertz@uni-bremen.de}{https://orcid.org/0000-0002-6347-1198}{}

\authorrunning{S. Braunfeld, J.\ Ne\v{s}e\v{r}il, P.\ Ossona de Mendez and S.\ Siebertz} 

\Copyright{Samuel Braunfeld, Jaroslav Ne\v{s}e\v{r}il, Patrice {Ossona de Mendez} and Sebastian Siebertz} 



\category{} 

\relatedversion{} 

\supplement{}

\funding{
	This paper is part of a project that has received funding from the European Research Council (ERC) under the European Union's Horizon 2020 research and innovation programme (grant agreement No 810115 -- {\sc Dynasnet}) and from the University of Bremen via Fokus project \emph{Graph classes with low twinwidth covers}. Samuel Braunfeld is further supported by Projects 21-10775S and 24-12591M of the Czech Science Foundation (GA\v{C}R).} 

\hideLIPIcs  

\EventEditors{Florin Manea and Alex Simpson}
\EventNoEds{2}
\EventLongTitle{30th EACSL Annual Conference on Computer Science Logic (CSL 2022)}
\EventShortTitle{CSL 2022}
\EventAcronym{CSL}
\EventYear{2022}
\EventDate{February 14--19, 2022}
\EventLocation{G\"{o}ttingen, Germany (Virtual Conference)}
\EventLogo{}
\SeriesVolume{216}
\ArticleNo{21}

\begin{document}
	
	\maketitle
	\begin{abstract}
			The notions of bounded-size and quasibounded-size decompositions with  bounded treedepth base classes are central to the structural theory of graph sparsity introduced by two of the authors years ago, and provide a characterization of both classes with bounded expansions and nowhere dense classes.

Strong connections of this theory with model theory led to considering  first-order transductions, which are logically defined graph transformations, and to initiate a comparative study of combinatorial and model theoretical properties of graph classes, with an emphasis on the model theoretical notions of dependence (or NIP) and stability.

In this paper, we first prove that the model theoretic notions of dependence and stability are, for hereditary classes of graphs, compatible with quasibounded-size decompositions, in the following sense: every hereditary class with quasibounded-size  decompositions with dependent (resp.\ stable)  base classes is itself dependent (resp.\ stable). This result is obtained in a more general study of ``decomposition horizons'', which are class properties compatible with quasibounded-size decompositions.

We deduce that hereditary classes with quasibounded-size decompositions with bounded shrubdepth base classes are stable. In the second part of the paper, we prove the converse. Thus, we characterize stable hereditary classes of graphs as those hereditary classes that admit quasibounded-size decompositions with bounded shrubdepth base classes. This result is obtained by proving that every hereditary stable class of graphs admits almost nowhere dense quasi-bush representations, thus answering positively a conjecture of Dreier et al.
 
 These results have several consequences. For example,
 we show that every graph $G$ in a stable, hereditary class of graphs $\mathscr C$ has a clique or a stable set of size $\Omega_{\mathscr C,\epsilon}(|G|^{1/2-\epsilon})$, for every $\epsilon>0$, which is tight in the sense that it cannot be improved to $\Omega_{\mathscr C}(|G|^{1/2})$.
	\end{abstract}
	\section{Introduction}

Initiated in 2005, the structural theory of graph sparsity \cite{Sparsity} aims to provide a general framework for the study of monotone classes of sparse graphs based on the notion of shallow minors. The notion of shallow minors was proposed by Leiserson and Toledo, and initially used by Plotkin, Rao, and Smith in their study of vertex separators \cite{plotkin1994shallow}. Recall that 
a class of graphs is \emph{monotone} if every subgraph of a graph in the class also belongs to the class, and that 
a \emph{depth-$r$ minor} of a graph $G$ is a graph that can be obtained from $G$ by contracting vertex-disjoint subgraphs with radius at most $r$ and deleting some vertices and edges. 

The notions of  a \emph{class with bounded expansion} \cite{POMNI} and of a \emph{nowhere dense class} \cite{ND_logic} are defined in terms of shallow minors: they are those classes of graphs whose depth-$r$ minors have average degree (resp.\ clique number) bounded by a function of $r$ (see formal definitions in Section~\ref{sec:prelims}). Note that these properties are preserved by taking subgraphs. Hence, a class~$\mathscr C$ has bounded expansion (resp.\ is nowhere dense) if and only if the same holds for the \emph{monotone closure} of $\mathscr C$ (that is, for the class of all the subgraphs of graphs in~$\mathscr C$).
 Some fundamental structural and algorithmic properties of these graph classes have been obtained using special decompositions, the so-called \emph{low treedepth decompositions} (also called \emph{low treedepth colorings})\footnote{Low treedepth decompositions/colorings correspond here to  bounded-size and quasibounded-size bounded treedepth decompositions.}
\cite{nevsetvril2008gradb,nevsetvril2011nowhere,Japan04}. 

As a generalization of these decompositions is central to this paper, we take time for introducing some definitions and terminology. A \emph{class property} $\Pi$ is a set of graph classes that is closed by subclasses (i.e.,  if $\mathscr C\in\Pi$ and $\mathscr D\subseteq\mathscr C$, then $\mathscr D\in\Pi$). For instance, ``(being) degenerate'', ``excluding a minor'', and ``(having) bounded maximum degree''  are class properties. 
For a function $f:\mathbb N\rightarrow\mathbb N$ and a parameter $p$, a class of graphs $\mathscr C$ has an 
\emph{$f$-bounded $\Pi$-decomposition with parameter $p$} if  there exists a graph class $\mathscr D_p\in\Pi$, such that  every graph $G\in\mathscr C$ has a vertex coloring by $f(|G|)$ colors with the property  that every $p$ color classes induce a subgraph in $\mathscr D_p$.  In this paper, we will consider only infinite classes of graphs and will be mainly interested in the asymptotic behavior of bounding functions. Particularly, we say that a class $\mathscr C$ has \emph{bounded-size} decompositions (resp.\ \emph{quasibounded-size} decompositions) if, for every integer $p$, it has an $f_p$-bounded $\Pi$-decomposition with parameter~$p$, where  $f_p(n)\in \mathcal O(1)$  (resp.\ $f_p(n)\in\mathcal O(n^\epsilon)$ for every $\epsilon>0$). Using this terminology, we have the following characterization theorems, based on the class property ``bounded treedepth'' (see Section~\ref{sec:prelims_gt} for a definition of treedepth).

\begin{theorem}[\cite{Taxi_stoc06,POMNI}]
	\label{thm:LTDBE}
	A  class $\mathscr C$ has bounded expansion if and only if it has bounded-size bounded treedepth decompositions.
\end{theorem}

\begin{theorem}[\cite{nevsetvril2011nowhere}]
	\label{thm:LTDND}
	A hereditary class $\mathscr C$ is nowhere dense if and only if it has quasibounded-size bounded treedepth decompositions.
\end{theorem}

Classes with bounded expansion and nowhere dense classes have strong algorithmic properties related to first-order logic \cite{DKT2,grohe2014deciding}, and it was conjectured that such a connection was ultimately due to the fact that, for monotone classes of graphs,  nowhere denseness is equivalent to the model theoretic notions of dependence and stability (as well as to their monadic versions) \cite{adler2014interpreting}; see Section~\ref{sec:prelims_mt} for a formal definition of these model theoretical notions. 
\pagebreak

\begin{theorem}[\cite{adler2014interpreting}]
	\label{thm:NDstable}
	For a monotone class of graphs $\mathscr C$, the following are equivalent:
	\begin{itemize}
		\item $\mathscr C$ is nowhere dense;
		\item $\mathscr C$ is stable;
		\item $\mathscr C$ is dependent;
		\item $\mathscr C$ is monadically stable;
\item $\mathscr C$ is monadically dependent.
	\end{itemize}
\end{theorem}

On the one hand, the notions of stability and dependence are central combinatorial properties in model theory, which delimit strong dividing lines between ``tame'' and ``wild'' theories. On the other hand, the model theoretical notions of interpretations and transductions (see Section~\ref{sec:prelims_trans}) suggested the possibility to study structural properties of graph classes, whose members are (uniformly) defined from a class of sparse graphs by using logical formulas. 
Unfortunately, stability and dependence are not in general preserved by transductions, while their monadic versions are. However, the collapse of these notions for monotone classes (as it appears in Theorem~\ref{thm:NDstable}) partially extends to hereditary classes.

\begin{theorem}[\cite{braunfeld2022existential}]
	\label{thm:LB}
	A hereditary class of graphs $\mathscr C$ is monadically dependent if and only if it is dependent; it is monadically stable if and only if it is stable.
\end{theorem}

This will allow us to interchangeably use ``stable'' and ``monadically stable'' and similarly ``dependent'' and ``monadically dependent'' when we speak about hereditary classes of graphs.

The notion of transduction led to the graph-theoretical concept of \emph{structural sparsity}~\cite{SurveyND}. For instance,  classes with bounded \emph{shrubdepth}, which are defined 
by the existence of very specific \emph{tree models}\footnote{A \emph{tree model} for a graph $G$ is a rooted tree $T$, whose set of leaves if $V(G)$, where every root-to-leaf path has length $d$ (the shrubdepth), every leaf gets one of $m$ labels ($m$~
	 being bounded in the class), and the adjacency of $u$ and $v$ in $G$ depends only on the labels  of $u,v$ and the distance between $u$ and $v$ in $T$.},
are characterized as those classes of graphs that are transductions of classes of bounded diameter trees~\cite{ganian2012trees}. They are a natural dense analog of classes with bounded treedepth, as witnessed by their  characterization as those classes of graphs having  bounded SC-depth~\cite{ganian2012trees} (see definition of SC-depth in Section~\ref{sec:Flipper}).
Classes with \emph{structurally bounded expansion} (i.e.,  transductions of classes with bounded expansion) have  been characterized in terms of bounded-size decompositions.

\begin{theorem}[\cite{SBE_drops}]
	A  class of graphs $\mathscr C$ has structurally bounded expansion if and only if it has bounded-size bounded shrubdepth decompositions.
\end{theorem}


Then, it has been conjectured that a similar characterization exists for structurally nowhere dense classes, that is, for transductions of nowhere dense classes. 

\begin{conjecture}
	\label{conj:SNDSD}
	A hereditary class of graphs $\mathscr C$ is structurally nowhere dense if and only if it has quasibounded-size bounded shrubdepth decompositions.
\end{conjecture}

However, the only published conjectures we have found mention only the left-to-right implication  \cite{kwon17} (see also \cite{brianski2021erdos,dreier2022treelike}), which has been proved recently.

\begin{theorem}[\cite{dreier2022treelike}]
	\label{thm:SNDSD}
	Every structurally nowhere dense class of graphs has quasibounded-size bounded shrubdepth decompositions.
\end{theorem}

On the other hand,  a natural extension of Theorem~\ref{thm:NDstable} to hereditary classes was conjectured.

\begin{conjecture}[{\cite[Conjecture 6.1]{nevsetril2020rankwidth}}]
	\label{conj:stSND}
	A hereditary class of graphs $\mathscr C$ is structurally nowhere dense if and only if it is (monadically) stable.
\end{conjecture}

Note that the left-to-right implication follows from Theorem~\ref{thm:NDstable}, as every structurally nowhere dense class is a transduction of a (monotone) nowhere dense class, and monadic stability is preserved under transductions.

Together with Conjectures~\ref{conj:SNDSD} and~\ref{conj:stSND},
it is natural to consider the following conjecture.
\begin{conjecture}
	\label{conj:stSD}
	A hereditary class of graphs $\mathscr C$ is stable if and only if it has quasibounded-size bounded shrubdepth decompositions.
\end{conjecture}
Indeed, any two of the Conjectures~\ref{conj:SNDSD}, \ref{conj:stSND}, and~\ref{conj:stSD}  imply the third. 

\subsection{Our Results}
In this paper we give a new characterization of hereditary stable classes that confirms Conjecture~\ref{conj:stSD}.

\setcounter{thmmain}{\value{theorem}}
\begin{theorem}
	\label{thm:stSD}
	A hereditary class of graphs $\mathscr C$ is stable if and only if it has quasibounded-size bounded shrubdepth decompositions.	
\end{theorem}

It may seem surprising at a first glance that general hereditary stable classes can be characterized in terms of decompositions into classes as restricted as classes with bounded shrubdepth.
This result is achieved by a combination of various recently introduced techniques (see in particular  \cite{dreier2022treelike,dreier2023firstorder}) together with a new general study 
 of bounded-size and quasibounded-size $\Pi$-decompositions, which  we present in Section~\ref{sec:horizons}. In this section, we focus particularly on class properties $\Pi$ so that every hereditary class with quasibounded-size $\Pi$-decompositions actually belongs to $\Pi$. We call such class properties \emph{decomposition horizons}, and we provide some examples of them, notably:

\begin{theorem}
	\label{thm:horizons}
	The class properties ``dependent'' and ``stable'' are decomposition horizons.
	In particular, hereditary classes with quasibounded-size bounded shrubdepth decompositions are stable.
\end{theorem}

The left-to-right implication of \cref{thm:stSD} is an extension of Theorem~\ref{thm:SNDSD}  based on a strengthening of the notion of neighborhood covers (see Section~\ref{sec:decst}).
Some consequences of Theorem~\ref{thm:stSD} are  discussed in Section~\ref{sec:main}. In particular, we show that it follows from  Theorem~\ref{thm:stSD} that every graph $G$ in a stable, hereditary class of graphs $\mathscr C$ contains a clique or a stable set of size
$\Omega_{\mathscr C,\epsilon}(|G|^{1/2-\epsilon})$ (for every $\epsilon>0$).

\medskip

This paper is an extended version of  the abstract  \emph{Decomposition horizons: from graph sparsity to model-theoretic dividing lines} \cite{covers_proc} presented at Eurocomb 2023. The main result of this paper, Theorem~\ref{thm:stSD}, as well as the results in  Sections~\ref{sec:decst} and~\ref{sec:main} are new.

\section{Preliminaries}\label{sec:prelims}
We assume basic familiarity with first-order logic and graph theory, and
refer e.g.\ to \cite{hodges1993model,diestel2012graph} for background
and for all undefined notation.  
All graphs considered here are undirected.
\subsection{Basic graph theory}
\label{sec:prelims_gt}
The vertex set of a graph $G$ is
denoted as~$V(G)$ and its edge set as $E(G)$. 
We denote by~$|G|$ the \emph{order} of $G$, that is, $|V(G)|$, and by $\|G\|$ its number of edges, that is, $|E(G)|$.
It will sometimes be convenient to use the adjacency relation~$E_G$ of $G$ instead of the edge set $E(G)$. Thus, $E_G(u,v)$ is another notation for $uv\in E(G)$.
All graphs considered
in this paper are finite. 
We denote by $K_t$ the complete graph (or \emph{clique}) on $t$
vertices and by~$K_{s,t}$ the complete bipartite graph (or \emph{biclique}) with parts of size~$s$ and $t$.

For a graph $G$ and a subset $A$ of vertices of $G$ we denote by $G[A]$ the subgraph of $G$ induced by $A$. We write $H\subseteq_i G$ if $H$ is an induced subgraph of $G$.
For a graph $G$ and subsets $A,B$ of vertices of $G$ we denote by $G[A,B]$ the subgraph of $G$ \emph{semi-induced} by $A$ and $B$, that is, the graph with  vertex set $A\cup B$, whose edges are those edges of $G$ with one endpoint in $A$ and the other endpoint in $B$. Note that $G[A]=G[A,A]$ and that $G[A,B]$ is bipartite if $A$ and $B$ are disjoint.

For a graph $G$ and an integer $p$, the $p$th subdivision $G^{(p)}$ of $G$ is the graph obtained by replacing each edge of $G$ by a path of length $p+1$. 
For two graphs $G$ and $H$, we denote by~$G+H$ the disjoint union of $G$ and $H$. 

A class of graphs $\mathscr C$ is \emph{monotone} (resp.\ \emph{hereditary}) if every subgraph (resp.\ induced subgraph) of a graph in $\mathscr{C}$ also belongs to $\mathscr{C}$.
Classes with bounded expansion and nowhere dense classes may be defined in several equivalent ways (see \cite{Sparsity}). 
The original definition is in terms of shallow minors.
The \emph{radius} of a connected graph $G$ is 
the minimum integer $r$ such that there exists a vertex $v\in V(G)$ at distance at most $r$ from all the other vertices of $G$.
A graph $H$ is a \emph{depth-$r$ minor} of a graph $G$ if there exists a mapping $\eta$ that maps vertices of $H$ to pairwise disjoint subgraphs of $G$, such that
\begin{itemize}
	\item for each $u\in V(U)$, $\eta(u)$ is connected and has radius at most $r$, 
	\item for each edge $uv$ of $H$, there is an edge of $G$ with one incidence in $\eta(u)$ and one incidence in $\eta(v)$.
\end{itemize}

A graph $H$ is a \emph{depth-$r$ topological minor} of $G$ if there exists an injective mapping 
\mbox{$\nu:V(H)\rightarrow V(G)$} and a family $\mathscr P$ of internally vertex-disjoint paths of $G$ with length at most $2r+1$ (with internal vertices not in the image of $\nu$), such that for every edge $uv\in E(H)$ there is a path $P_{uv}\in\mathscr P$ with endpoints $\nu(u)$ and $\nu(v)$.
A class $\mathscr C$ is \emph{nowhere dense} if, for every integer $r$, the clique number of the depth-$r$ minors of the graphs in $\mathscr C$ is bounded by a constant $C(r)$; it has \emph{bounded expansion} if, for every integer $r$, the average degree of the depth-$r$ minors of the graphs in $\mathscr C$ is bounded by a constant $D(r)$.

In this paper, it will be convenient to use the following characteristic properties (cf.\  \cite{Sparsity}).
A class $\mathscr C$ has \emph{bounded expansion} if there exists a function $f:\mathbb N\rightarrow\mathbb N$ with the following property:
if $H$ is a graph with the property that $H^{(p)}$ is a subgraph of a graph in $\mathscr C$, then the average degree of $H$ is at most $f(p)$.
A class $\mathscr C$ is \emph{nowhere dense} if there exists a function $f:\mathbb N\rightarrow\mathbb N$ with the following property:
if $H$ is a graph with the property that $H^{(p)}$ is a subgraph of a graph in $\mathscr C$, then the clique number  of $H$ is at most~$f(p)$.

We denote by $\nabla_r(G)$ the \emph{rank-$r$ greatest average density} of $G$, that is, the maximum  edge density of the depth-$r$ minors of $G$:
\[
\nabla_r(G)=\max\biggl\{\frac{|E(H)|}{|V(H)|}\colon H\text{ is a depth-$r$ minor of $G$}\biggr\}.
\]

Similarly, we denote by $\widetilde{\nabla}_r(G)$ the \emph{rank-$r$ topological greatest average density} of $G$, that is, the maximum  edge density of the depth-$r$ topological minors of $G$:
\[
\widetilde{\nabla}_r(G)=\max\biggl\{\frac{|E(H)|}{|V(H)|}\colon H\text{ is a depth-$r$ topological minor of $G$}\biggr\}.
\]

It is clear from the definition that a class has bounded expansion if
$\nabla_r(G)\in \mathcal O_{\mathscr C,r}(1)$ holds for graphs $G\in\mathscr C$. 
It is known \cite{Sparsity} that this is equivalent to $\widetilde{\nabla}_r(G)\in\mathcal O_{\mathscr C,r}(1)$.
We can also use the greatest average densities to characterize nowhere-denseness.

\begin{theorem}[\cite{nevsetvril2011nowhere}]
	For a  hereditary class $\mathscr C$ of graphs, the following are equivalent:
	\begin{enumerate}
\item $\mathscr C$  is nowhere dense;
\item $\nabla_r(G)\in \mathcal O_{\mathscr C,r,\epsilon}(|G|^\epsilon)$, for every $r\in\mathbb N$, every $\epsilon>0$, and graph $G\in \mathscr C$.
\item $\widetilde{\nabla}_r(G)\in\mathcal O_{\mathscr C,r,\epsilon}(|G|^\epsilon)$, for every $r\in\mathbb N$, every $\epsilon>0$, and graph $G\in \mathscr C$.
	\end{enumerate}
	\end{theorem}

Recently, a weakening of the notion of nowhere-denseness has been proposed, in terms of bounding of weak coloring numbers~\cite{dreier2022treelike}.
 As weak coloring numbers are polynomially equivalent to the greatest average densities, we can reformulate the definition as follows: 
A class of graphs $\mathscr C$ is \emph{almost nowhere dense} if $\nabla_r(G)\in\mathcal O_{\mathscr C,r,\epsilon}(|G|^\epsilon)$, for every $r\in\mathbb N$ and every $\epsilon>0$ or, equivalently, $\widetilde{\nabla}_r(G)\in\mathcal O_{\mathscr C,r,\epsilon}(|G|^\epsilon)$, for every $r\in\mathbb N$ and every $\epsilon>0$. Note that an almost nowhere dense class is not assumed to be hereditary, and that a hereditary almost nowhere dense class  is nowhere dense.

The {\em treedepth} of a graph is a minor montone graph
invariant that has been defined in~\cite{Taxi_tdepth}, and is
	defined as
	the minimum height of a rooted forest $Y$ such that $G$ is a subgraph of the closure of $Y$ (that is, of the graph obtained by adding edges between a vertex and all its ancestors). 
The treedepth of a graph $G$ can be defined inductively by
${\rm td}(K_1)=1$, ${\rm td}(G+H)=\max({\rm td}(G),{\rm td}(H))$, and 
${\rm td}(G)=1+\min_{v\in V(G)}{\rm td}(G-v)$ if $G$ is connected and $G\not\simeq K_1$.

\subsection{Basic model theory}
\label{sec:prelims_mt}
In this paper we consider either graphs or {\em $\sigma$-expanded
	graphs}, that is, graphs with additional unary relations in $\sigma$
(for a set $\sigma$ of unary relation symbols). 
In particular, a $\sigma$-expanded graph obtained by adding to a graph $G$ additional unary relations in $\sigma$ will be called a \emph{$\sigma$-expansion} of $G$.
We usually denote
graphs by $G,H,\ldots$ and $\sigma$-expanded graphs by
$G^+,H^+,G^*,H^*,\ldots$,
but sometimes we will use $G,H,\ldots$ for
$\sigma$-expanded graphs as well. 
We shall often use the term ``colored graph'' instead of $\sigma$-expanded graph.
By extension, a \emph{$\sigma$-expansion} of  a class~$\mathscr C$ is a class containing, for each $G\in\mathscr C$, a $\sigma$-expansion of $G$.
In formulas, the adjacency relation
will be denoted as $E(x,y)$. Whenever we speak of formulas we will mean first-order formulas. For a formula $\phi(\bar x)$ and a (colored) graph $G$, we denote by
$\phi(G)$ the set all tuples $\bar v\in V(G)^{|\bar x|}$ such that  $G$ satisfies $\phi(\bar v)$.

\medskip

We recall some classical model-theoretical notions due to Shelah.
\medskip

A class $\mathscr C$ is \emph{unstable} if there exists a formula $\varphi(\bar x, \bar y)$ such that, for every $n\in \N$ there exist a graph $G\in\mathscr C$ and $2n$ tuples of vertices
$\bar u_1,\dots,\bar u_n\in V(G)^{|\bar x|}$ and $\bar v_1,\dots,\bar v_n\in V(G)^{|\bar y|}$ with
\[
G\models\varphi(\bar u_i,\bar v_j)\qquad\iff\qquad i\leq j.
\]

If $\mathscr{C}$ is not unstable, then it is \emph{stable}. $\mathscr C$ is \emph{monadically stable} if every $\sigma$-expansion of $\mathscr C$ is stable. 

A class $\mathscr C$ is \emph{independent} if there exists a formula $\varphi(\bar x, \bar y)$ such that, for every $n\in \N$ there exist a graph $G\in\mathscr C$ and $n+2^n$ tuples of vertices
$\bar u_i\in V(G)^{|\bar x|}$ ($i\in [n]$) and $\bar v_J\in V(G)^{|\bar y|}$ ($J\subseteq [n]$) with
\[
G\models\varphi(\bar u_i,\bar v_J)\qquad\iff\qquad i\in J.
\]

If $\mathscr{C}$ is not independent, then it is \emph{dependent} (or \emph{NIP}). $\mathscr C$ is \emph{monadically dependent} if every $\sigma$-expansion of $\mathscr C$ is dependent.

\subsection{Transductions}\label{sec:transductions}
\label{sec:prelims_trans}
For a set $\sigma$ of unary relations, the \emph{coloring operation} $\Gamma_\sigma$ maps a graph $G$ to the set~$\Gamma_\sigma(G)$ of all its $\sigma$-expansions.

A \emph{simple  interpretation} $\mathsf I$ of graphs in \mbox{$\sigma$-expanded}
graphs is a pair $(\nu(x), \eta(x,y))$ consisting of two formulas (in
the first-order language of $\sigma$-expanded graphs), where $\eta$ is symmetric
and anti-reflexive (i.e.,
\mbox{$\models \eta(x,y)\leftrightarrow\eta(y,x)$} and
\mbox{$\models \eta(x,y)\rightarrow\neg(x=y)$}). If $G^+$ is a
$\sigma$-expanded graph, then $H=\mathsf I(G^+)$ is the graph with
vertex set $V(H)=\nu(G^+)$ and edge set $E(H)=\eta(G^+)\cap \nu(G^+)^2$.

A \emph{transduction}\footnote{In this paper, we don't consider the possibility for a transduction to copy vertices; what we call transduction here is called a \emph{non-copying transduction} in \cite{braunfeld2022first}.} $\mathsf T$ is the composition
$\mathsf I\circ\Gamma_\sigma$ of a  coloring operation~$\Gamma_\sigma$ and a simple interpretation $\mathsf I$ of graphs in $\sigma$-expanded graphs.
In other words, for every graph $G$ we have
$\mathsf T(G)=\{\mathsf I(H^+): H^+\in\Gamma_\sigma(G)\}$.
We denote by $\mathsf{Id}$ the \emph{identity transduction}, which is such that $\mathsf{Id}(\mathbf A)=\{\mathbf A\}$ for every structure $\mathbf A$.
The transduction $\mathsf T$ is \emph{quantifier-free} if all the formulas defining the interpretation part of $\mathsf T$ are quantifier-free.
A {\em subset complementation} transduction is defined by the
quantifier-free interpretation on a $\sigma$-expansion (with
$\sigma=\{M\}$) by
$\eta(x,y):=(x\neq y)\wedge \neg \bigl(E(x,y)\leftrightarrow(M(x)\wedge M(y)\bigr)$.
In other words, the subset complementation transduction complements
the adjacency inside the subset of the vertex set defined by~$M$. We
denote by $\comp M$ the subset complementation defined by the unary
relation~$M$. A {\em perturbation} is a composition of (a bounded
number of) subset complementations.
For a class $\Dd$ and a transduction $\mathsf T$ we define $\mathsf T(\Dd)=\bigcup_{G\in\Dd}\mathsf T(G)$ and 
we say that a class~$\Cc$ is a  {\em $\mathsf T$-transduction} of $\Dd$ if $\Cc\subseteq\mathsf T(\Dd)$. 

Transductions allow for an alternative characterization of monadic stability and monadic dependence, using results of \cite{baldwin1985second}. We first recall that the {\em half-graph of length $n$} is the bipartite graph with vertices $\{a_i, b_i : i \in [n]\}$ such that $E(a_i, b_j) \iff i \leq j$.

\begin{proposition}
	A class $\mathscr C$ is monadically dependent if and only if it does not transduce the class of all finite graphs. 
	
	A class $\mathscr C$ is monadically stable if and only if it does not transduce the class of all finite half-graphs.
\end{proposition}

%
%

Transductions are at the heart of the definition of shrubdepth, which is a measure of the structural complexity of graph classes (and not of single graphs).
A class $\mathscr D$ has \emph{shrubdepth} at most $d$ if it is a transduction of a class of forests with radius at most $d$ \cite{ganian2012trees}. A more combinatorial characterization of bounded shrubdepth classes, in terms of SC-depth, reflects the proximity to cographs (see Section~\ref{sec:Flipper}).

\section{Decomposition horizons}
\label{sec:horizons}
\subsection{$\Pi$-decompositions}
Decompositions aim to break complex graphs into simpler ones. 
In this section, we generalize the decompositions used in \cref{thm:LTDBE} and \cref{thm:LTDND} and establish some of their abstract properties.
We define a \emph{hereditary class property} to be a downset $\Pi$ of hereditary graph classes, that is, a set of hereditary classes such that if $\mathscr C\in\Pi$ and $\mathscr D$ is a hereditary class with $\mathscr D\subseteq \mathscr C$, then $\mathscr D\in\Pi$.

\begin{definition}
\label{def:decomp}
Let $\Pi$ be a hereditary class property, let $f:\mathbb N\rightarrow\mathbb N$ be a non-decreasing function and let $p$ be a positive integer.
We say that a class $\mathscr C$ has an \emph{$f$-bounded  \mbox{$\Pi$-decomposition}} with parameter $p$ if there exists $\mathscr D_p\in\Pi$ such that, for every graph $G\in\mathscr C$, there exists an integer $N\leq f(|G|)$ and a partition $V_1,\dots,V_N$ of the vertex set of $G$
with $G[V_{i_1}\cup\dots\cup V_{i_p}]\in \mathscr D_p$ for all $i_1,\dots,i_p\in [N]$.

When $f$ is a bounded function, we say that $\mathscr C$ has a  \emph{bounded-size $\Pi$-decomposition} with parameter $p$; when $f$ is a function with $f(n)=\mathcal O_\epsilon(n^{\epsilon})$ for every $\epsilon>0$, we say that $\mathscr C$ has a \emph{quasibounded-size $\Pi$-decomposition} with parameter $p$. If a class $\mathscr{C}$ has a bounded-size (resp.\ a quasibounded-size) $\Pi$-decomposition with parameter $p$ for each positive integer $p$, we say that $\mathscr C$ has \emph{bounded-size $\Pi$-decompositions} (resp.\ \emph{quasibounded-size $\Pi$-decompositions}). 
\end{definition}

For instance, by \cref{thm:LTDBE} and \cref{thm:LTDND},  considering the hereditary class property  ``bounded treedepth'', 
we have that a class $\mathscr C$ has bounded-size bounded treedepth decompositions if and only if it has bounded expansion, and it has quasibounded-size bounded treedepth decompositions if and only if it is nowhere dense. 

We can interpret these definitions as operators, constructing a (possibly larger) class property by the existence of special decompositions into a given base class property. This can be formulated as follows:
\begin{definition}
For a hereditary class property $\Pi$ we define the properties $\Pi^{+}$ (resp.\ $\Pi^{\ast}$) as follows:
\begin{itemize}
	\item $\mathscr C\in \Pi^{+}$ if $\mathscr C$ has bounded-size $\Pi$-decompositions;
	\item $\mathscr C\in \Pi^{\ast}$ if $\mathscr C$ has quasibounded-size $\Pi$-decompositions.
\end{itemize}
\end{definition}
In this notation, for $\Pi=$ `bounded treepdepth', we have $\Pi^+=$ `bounded expansion' and $\Pi^\ast=$ `nowhere dense'.

Note that $\Pi\subseteq\Pi^+\subseteq\Pi^\ast$ and that $\Lambda\subseteq\Pi$ implies both
$\Lambda^+\subseteq \Pi^+$ and $\Lambda^\ast\subseteq \Pi^\ast$. Also note that $\Pi^+$ and $\Pi^*$ are hereditary class properties. 
The next lemma shows that $\Pi\mapsto\Pi^+$ is a closure operator.

\begin{lemma}
Let $\Pi$ be a hereditary class property, let $f,g$ be non-decreasing functions, let $p$ be a non-negative integer, let~$\Lambda$ be the hereditary class property ``has an $f$-bounded $\Pi$-decomposition with parameter $p$'', and let $\Upsilon$ be the hereditary class property ``has a $g$-bounded $\Lambda$-decomposition with parameter $p$''.

Then every class $\mathscr C\in\Upsilon$ has an $h$-bounded $\Pi$-decomposition with parameter $p$, where 
\[
h(n)=g(n)\,f(n)^{\binom{g(n)-1}{p-1}}.
\]
\end{lemma}
\begin{proof}
Let $\mathscr C\in \Upsilon$. By definition, for every integer $p$ there exists a class $\mathscr D_{\Lambda,p}\in\Lambda$ with the property that every graph $G\in\mathscr C$ has a vertex partition into $g(|G|)$ parts $V_1(G),\dots,V_{g(|G|)}(G)$, such that every  $p$ parts induce a subgraph in $\mathscr D_{\Lambda,p}$.

Similarly, as $\mathscr D_{\Lambda,p}\in \Lambda$,
 there exists  a class $\mathscr D_{\Pi,p}\in\Pi$  with the property that every graph $H\in\mathscr D_{\Lambda,p}$ has a vertex partition into $f(|H|)$ parts $W_1(H),\dots,W_{f(|H|)}(H)$, such that every  $p$ parts  induce a subgraph in $\mathscr D_{\Pi,p}$. 

Let $G\in\mathscr C$, let $N_\Lambda=g(|G|)$, and let $N_\Pi=f(|G|)$.

For each subset $I$ of $[N_\Lambda]$ of size~$p$, denote by $G_I$ the subgraph of $G$ induced by the union of the parts $V_i(G)$ with $i\in I$. As $f$ is non-decreasing, 
$f(|G_I|)\leq f(|G|)=N_\Pi$. 
As above, we denote by 
$W_1(G_I),\dots,W_{N_\Pi}(G_I)$ the parts of the $\Pi$-decomposition of $G_I$ (with possibly empty last parts if $N_\Pi>f(|G_I|)$).

We derive a partition of $V(G)$ such that every $p$ parts induce a subgraph in $\mathscr D_{\Pi,p}$.
For  $i\in [N_\Lambda]$, let $\mathcal I_i$ be the set of all
subset $I$ of $[N_\Lambda]$ of size $p$ that contain $i$, and let $\mathcal F_i$ of  all the  functions from $\mathcal I_i$ to $[N_\Pi]$.
The partition of $V(G)$ we construct has  parts $Z_{i,\zeta}$ indexed by pairs $(i,\zeta)$, where  $i\in [N_\Lambda]$, and $\zeta\in\mathcal F_i$.

Intuitively, the part of a vertex $v\in V(G)$ will be defined by the part $V_i(G)$ it belongs to (in the partition of $V(G)$) as well as the indication, for each subset of $[N_\Lambda]$ of size $p$ containing the index $i$ of $v$, of the part $W_{\zeta(I)}$ of $G_I$ that contains $v$.
Formally, for $i\in[N_\Lambda]$ and $\zeta\in \mathcal F_i$ we define
\[
Z_{i,\zeta}=V_i\cap\bigcap_{I\in\mathcal I_i}W_{\zeta(I)}(G_I).
\]

It is easily checked that the sets $Z_{i,\zeta}$ partition $V(G)$.  We now prove that every $p$ parts induce a subgraph in $\mathscr D_{\Pi,p}$:
Let $i_1,\dots,i_p\in [N_\Lambda]$,
 let $\zeta_1\in\mathcal F_1,\dots,\zeta_p\in\mathcal F_p$, and  let $J$ be any  subset of $[N_\Lambda]$  of size $p$ that includes $i_1,\dots,i_p$. Note that $J\in \mathcal I_j$ for all $1\leq j\leq p$. It follows that
 $Z_{i_j,\zeta_j}=V_{i_j}\cap\bigcap_{I\in\mathcal I_{i_j}}W_{\zeta_j(I)}(G_I)\subseteq W_{\zeta_j}(G_J)$.
Hence,  
 $G[Z_{i_1,\zeta_1}\cup\dots\cup Z_{i_p,\zeta_p}]$ is an induced subgraph
of $G[W_{\zeta_1(J)}(G_J)\cup\dots\cup W_{\zeta_p(J)}(G_J)]=G_J[W_{\zeta_1(J)}(G_J)\cup\dots\cup W_{\zeta_p(J)}(G_J)]$.

Hence, $G[Z_{i_1,\zeta_1}\cup\dots\cup Z_{i_p,\zeta_p}]\in\mathscr D_{\Pi,p}$. It follows that $G$ has a  $h$-bounded $\Pi$-decomposition with parameter $p$, where $h(n)=g(n)\,f(n)^{\binom{g(n)-1}{p-1}}$.
\end{proof}

Considering the case where $g(n)=\mathcal{O}(1)$  and where $f(n)$ is either $\mathcal{O}(1)$  or $\mathcal O_\epsilon(n^{\epsilon})$ for every $\epsilon>0$, we get the following.
\begin{corollary}
	\label{cor:idem}
For every hereditary class property $\Pi$, we have $(\Pi^+)^+=\Pi^+$ and $(\Pi^\ast)^+=\Pi^\ast$.
\end{corollary}

\begin{lemma}
\label{lem:inter}
Let $\mathscr C$ have an $f_i$-bounded $\Pi_i$-decomposition with parameter $p$, for $i=1,2$.
Then $\mathscr C$ has an $f_1f_2$-bounded $\Pi_1\cap\Pi_2$-decomposition with parameter $p$.
\end{lemma}
\begin{proof}
By assumption, for each $i\in\{1,2\}$, there exist $\mathscr D_i\in\Pi_i$ such that every graph $G\in\mathscr C$ has a vertex partition into $N_i\leq f_i(|G|)$ parts $V_{i,1},\dots,V_{i,N_i}$, each $p$ parts inducing a graph in $\mathscr D_i$.
Consider the partition of the vertex set of $G$ into the sets $W_{i,j}=V_{1,i}\cap V_{2,j}$.
Then, any $p$ parts $W_{i_1,j_1},\dots,W_{i_p,j_p}$ induce a subgraph $H$, which is an induced subgraph of both $G[\bigcup_{k=1}^p V_{1,i_k}]$ and $G[\bigcup_{k=1}^p V_{2,j_k}]$, thus $H\in\mathscr D_1\cap\mathscr D_2$. It follows that $\mathscr C$ has an $f_1f_2$-bounded $\Pi$-decomposition with parameter $p$.
\end{proof}
As $(\Pi_1\cap\Pi_2)^+\subseteq\Pi_1^+\cap \Pi_2^+$ and $(\Pi_1\cap\Pi_2)^\ast\subseteq\Pi_1^\ast\cap \Pi_2^\ast$, we deduce
\begin{corollary}
\label{cor:int}
For every two hereditary class properties $\Pi_1$ and $\Pi_2$, we have
\[
(\Pi_1\cap\Pi_2)^+=\Pi_1^+\cap \Pi_2^+\qquad\text{and}\qquad(\Pi_1\cap\Pi_2)^\ast=\Pi_1^\ast\cap \Pi_2^\ast
\]
In particular,
$\Pi=\Lambda_1^+=\Lambda_2^+$ implies $ \Pi=(\Lambda_1\cap\Lambda_2)^+$ and
$\Pi=\Lambda_1^\ast=\Lambda_2^\ast$ implies $\Pi=(\Lambda_1\cap\Lambda_2)^\ast$.
\end{corollary}

\begin{lemma}
\label{rem:union}
Let $(\Pi_i)_{i\in I}$ be a family of hereditary class properties indexed by a (possibly infinite) set $I$.
Then $\mathscr C$ has an $f$-bounded $\bigcup_{i\in I}\Pi_i$-decomposition with parameter $p$ if and only if $\mathscr C$  has an $f$-bounded $\Pi_i$-decomposition with parameter $p$ for some $i\in I$.

In particular, $(\bigcup_{i\in I}\Pi_i)^+=\bigcup_{i\in I}\Pi_i^+$ and
$(\bigcup_{i\in I}\Pi_i)^\ast=\bigcup_{i\in I}\Pi_i^\ast$.
\end{lemma}
\begin{proof}
Assume $\mathscr C$ has an $f$-bounded  $\bigcup_{i\in I}\Pi_i$-decomposition with parameter $p$ .
Then, there exists $\mathscr D\in \bigcup_{i\in I}\Pi_i$ such that every graph $G\in\mathscr C$ has a vertex partition into $f(|G|)$ parts, any $p$ of them inducing a subgraph in $\mathscr D$.
Thus, for some $i\in I$ we have  $\mathscr D\in\Pi_i$, hence $\mathscr C$  has an $f$-bounded $\Pi_i$-decomposition with parameter $p$.

Conversely, if $\mathscr C$  has an $f$-bounded $\Pi_i$-decomposition with parameter $p$ for some $i\in I$, then it obviously has an $f$-bounded  $\bigcup_{i\in I}\Pi_i$-decomposition with parameter $p$.
\end{proof}

\begin{definition}
	We say that a hereditary class property $\Pi$ is a \emph{decomposition horizon} if $\Pi^\ast=\Pi$.  
	If $\Lambda$ is a hereditary class property, the \emph{decomposition horizon of}  $\Lambda$ is the 
	smallest decomposition horizon including $\Lambda$.
\end{definition}	

According to \Cref{rem:union},  $\bigcup_{k\geq 1}\Lambda^{\overbrace{\ast\dots\ast}^{k\text{ times}}}$ is a decomposition horizon, which is obviously the smallest decomposition horizon including $\Lambda$. Thus, the decomposition horizon of a hereditary class property is well-defined.
Also note that if $\Pi$ is a decomposition horizon, then $\Pi^+=\Pi$.

Recall that a topology on a set $X$ is a collection $\mathcal T$ of subsets of $X$ with the following properties. 1) $\emptyset\in \mathcal T$ and $X\in \mathcal T$, 2) $\mathcal T$ is closed under arbitrary unions, and 3) $\mathcal T$ is closed under finite intersections. 
As both $\emptyset$ and the set of all hereditary classes are hereditary class properties, 
the next result is  a direct consequence of \Cref{cor:int,rem:union}.
\begin{corollary}
Each of 
the set of all hereditary properties $\Pi$ with $\Pi^+=\Pi$ and 
the set of all decompositions horizons 
defines a topology on the set of all hereditary graph classes.
\end{corollary}

\begin{example}
\label{ex:forb}
For  a graph $H$, let $\Phi_H$ be the hereditary class property of all hereditary classes excluding $H$.
Then, for every graph $H$, the hereditary class property $\Phi_H$ is a decomposition horizon.
\end{example}
\begin{proof}
Assume a hereditary class $\mathscr C$ has $f$-bounded $\Phi_H$-decompositions, for some non-decreasing function  $f:\mathbb N\rightarrow\mathbb N$.
Clearly, $H\notin\mathscr C$ (consider parameter $p=|H|$), thus $\mathscr C\in\Phi_H$.
\end{proof}

We remark that the above example extends for class properties defined by excluding finitely many graphs, according to \cref{cor:int}. \cref{ex:forb} can also be extended as follows:
\begin{example}
Let $H$ be a graph and let $\Pi$ be the hereditary class property of classes $\mathscr D$ such that 
\[
\limsup_{G\in\mathscr D}\frac{\log \#H\subseteq_i G}{\log |G|}\leq \alpha,
\]
where $\# H\subseteq_i G$ denotes the number of induced subgraphs of $G$ isomorphic to $H$.
Then, $\Pi$ is a decomposition horizon.
\end{example}
\begin{proof}
	First notice that either $\alpha\geq 0$ or $\Pi$ is the hereditary class property of being \mbox{$H$-free}, which we considered in Example~\ref{ex:forb}. 
	Hence, we can assume $\alpha\geq 0$.
Assume $\mathscr C$ has quasibounded $\Pi$-decompositions. Then, for $p=|H|$, there exists a function $f$ with $f(n)\in\mathcal O_\epsilon(n^{\epsilon})$ for every $\epsilon>0$ and a hereditary class $\mathscr D_p\in\Pi$ such that for every graph $G\in\mathscr C$, the vertex set of $G$ has a partition into $N=f(|G|)$ parts $V_1,\dots,V_N$, any $p$ of them inducing a graph in $\mathscr D_p$.
Obviously, 
\[
\#H\subseteq_i G\leq\sum_{I\in\binom{[N]}{p}}(\#H\subseteq_i G_I)\leq N^p\,\max _{I\in\binom{[N]}{p}}(\#H\subseteq_i G_I).
\]
Thus, as $N^p=\mathcal O_\epsilon(|G|^{\epsilon})$ for every $\epsilon>0$ (apply $N\in \mathcal O_\epsilon(n^\epsilon)$ to $\epsilon/p$), we have 

\begin{align*}
N^p \,\max _{I\in\binom{[N]}{p}}(\#H\subseteq_i G_I)&\leq N^p \,\max _{I\in\binom{[N]}{p}}|G_I|^\alpha & \text{(as $G_I\in\mathscr D_p$)}\\
&\leq N^p\, |G|^\alpha &\text{(as $|G_I|\leq |G|$ and $\alpha\geq0$)}\\
&\in \mathcal O_\epsilon(|G|^{\alpha+\epsilon}).
\end{align*}
Hence,
\[
\limsup_{G'\in\mathscr C}\frac{\log \#H\subseteq_i G'}{\log |G'|}\leq \alpha.
\]
Hence $\mathscr C\in\Pi$. It follows that $\Pi^\ast=\Pi$, that is, $\Pi$ is a decomposition horizon. 
\end{proof}

The following immediate lemma shows that $\Pi$-decompositions behave nicely with respect to quantifier-free transductions.

\begin{lemma}
\label{lem:QF}
Let $\mathsf T$ be a quantifier-free transduction. If a class $\mathscr C$ has an $f$-bounded  $\Pi$-decomposition with parameter $p$, then $\mathsf T(\mathscr C)$ has an $f$-bounded  $\mathsf T(\Pi)$-decomposition with parameter $p$, where $\mathsf T(\Pi)=\{\mathscr D\colon \exists\mathscr D'\in\Pi\text{ with }\mathscr D\subseteq\mathsf T(\mathscr D')\}$.
\end{lemma}

\begin{corollary}
\label{cor:QF}
For every quantifier-free transduction $\mathsf T$ we have
$\mathsf T(\Pi^+)\subseteq \mathsf T(\Pi)^+$ and $\mathsf T(\Pi^\ast)\subseteq \mathsf T(\Pi)^\ast$. 

For every perturbation  $\mathsf P$ we have
$\mathsf P(\Pi^+)=\mathsf P(\Pi)^+$ and $\mathsf P(\Pi^\ast)= \mathsf P(\Pi)^\ast$. 
In particular, if~$\Pi$ is a decomposition horizon, then $\mathsf P(\Pi)$ is a decomposition horizon.

Also, if $\Pi$ is a decomposition horizon, then the hereditary class property ``perturbation of a class in $\Pi$'' is a decomposition horizon.
\end{corollary}
\begin{proof}
The second property follows from the easily checked property of perturbations  that if $G'\in\mathsf P(G)$, then $G\in\mathsf P(G')$. The last property follows from \Cref{rem:union}, by considering all possible perturbations.
\end{proof}

Note that \cref{lem:QF} extends to general structures. However, in this latter case, there can be some shift in the value of the parameter.

\subsection{Some decomposition horizons}
In this part, we prove that the several hereditary class properties are decomposition horizons.
Some have a clear graph-theoretic definition, like
$\Delta$ (bounded maximum degree), $\Delta_k$~(bounded maximum degree after deletion of at most $k$ vertices), $\Delta_\omega$ (bounded maximum degree after deletion of a bounded number of vertices), $\Sigma$ (weakly sparse), and nowhere dense;
some have more model-theoretic significance: mutually algebraic, stable, and dependent.
That the hereditary class properties ``stable'' and ``dependent'' are decomposition horizons is one of the main results of this paper.

\begin{theorem}
\label{thm:delta}
The class property $\Delta=$ ``bounded maximum degree'' is the decomposition horizon of $\Lambda=$``connected components of bounded size''. Moreover, $\Lambda^\ast=\Lambda^+=\Delta$. 
\end{theorem}
\begin{proof}
Assume that the maximum degree of the graphs in a class $\mathscr C \in \Delta^*$ is unbounded. Then, the class contains, for every integer $n$, a graph $G_n$ with $n$ vertices, one of which is connected to all the others. In any $N$-coloring of $G_n$, some subgraph induced by two colors is a connected subgraph with maximum degree at least $(n-1)/N$, which is a contradiction. In particular, fix $\epsilon < 1$, let $f(n) \leq cn^\epsilon$ for some $c \in \mathbb R$ be such that $\mathscr C$ admits $f$-bounded $\Delta$-decompositions with parameter 2, and let $N = f(n)$. Then $(n-1)/N \geq (n-1)/(cn^\epsilon)$ goes to infinity with $n$. Thus, $\Delta^\ast\subseteq\Delta$, which means that $\Delta$ is a decomposition horizon.

We first show $\Delta\subseteq\Lambda^+$: 
Let $\mathscr C\in\Delta$ and let $d$ be the maximum degree of the graphs in~$\mathscr C$.
Fix a positive integer $p$. Let $G^p$ be the $p$th power of $G$, that is, the graph obtained from $G$ by placing an edge between any two points at distance at most $p$. For every graph $G\in\mathscr C$, consider a proper coloring of $G^p$ with $d^p+1$ colors. Then, every path of length $p$ of $G$ gets at least $p+1$ colors. Thus, every $p$ classes induce a graph in $\Lambda$. Thus, $\Delta\subseteq\Lambda^+$. 
Moreover, as $\Lambda\subseteq\Delta$ we have $\Lambda^\ast\subseteq\Delta^\ast=\Delta$.
Thus, $\Delta\subseteq\Lambda^+\subseteq\Lambda^\ast\subseteq\Delta$.
\end{proof}

From this it will follow that
``bounded maximum degree after deletion of $k$ vertices'', and
``bounded maximum degree after deletion of a bounded number of vertices'' are decomposition horizons.

\begin{theorem}
Let $k$ be a non-negative integer.
The class property $\Delta_k=$ ``bounded maximum degree after deletion of at most $k$ vertices'' is a decomposition horizon. 
\end{theorem}
\begin{proof}
Assume that the maximum degree of the graphs in a class $\mathscr C$ remains unbounded after deletion of $k$ vertices. Then, the class contains, for every integer $n$, 
a graph $H_n$ with at least $k+1$ vertices with degree at least $n$.
By taking an induced subgraph, we get a graph~$G_n$ with at most $(k+1)(n+1)$ vertices, including $k$ vertices with degree at least $n$.  In any $N$-coloring of $G_n$, some subgraph induced by $2(k+1)$ colors is a connected subgraph with at least $k+1$ vertices with degree  at least $n/N$. Thus, $\Delta_k^\ast\subseteq\Delta_k$, which means that $\Delta_k$ is a decomposition horizon.
\end{proof}
\begin{corollary} \label{cor:nearly bounded degree}
The class property $\Delta_\omega=$ ``bounded maximum degree after deletion of a bounded number of vertices'' is a decomposition horizon. 
\end{corollary}
\begin{proof}
Let $\Delta_k=$ ``bounded maximum degree after deletion of at most $k$ vertices''. Then,	$\Delta_\omega=\bigcup_{k\geq 0}\Delta_k$.  So, the result follows from \Cref{rem:union}.
\end{proof}

\begin{theorem}
The class property ``transduction of a class with bounded maximum degree''  is a decomposition horizon.
\end{theorem}
\begin{proof}
As noted in \cite{GajarskyHOLR16},  a class $\mathscr{C}$ is a transduction of a class with bounded maximum degree if and only if it is a perturbation of  a class with bounded maximum degree. We conclude with \cref{thm:delta} and  \cref{cor:QF}.
\end{proof}

Also, it is noted in \cite{braunfeld2022first} that ``transduction of a class with bounded maximum degree'' is equivalent to the model-theoretic property of ``mutually algebraic'', which \cite{laskowski2013mutually} in turn shows is equivalent to the model-theoretic property ``monadic NFCP''.

\b
Recall that a class is \emph{weakly sparse} (or \emph{biclique-free}) if, for some integer $t$, no graph in the class contain $K_{t,t}$ as a subgraph. 
One  of the most famous results in extremal combinatorics provides 
an upper bound of the maximum number of edges ${\rm ex}(n,K_{s,t})$  of a graph with~$n$~vertices that does not contain $K_{s,t}$.
\def\xxx{K{\H o}v\'ari--S\'os--Tur\'an  \cite{KST}}
\begin{theorem}[\xxx]
\label{thm:KST}
For every two integers $t\geq s\geq 2$, 
\[
{\rm ex}(n,K_{s,t})\leq\frac12(t-1)^{\frac1s}n^{2-\frac1s}+\frac12(s-1)n.
\]
In particular\footnote{(as $\frac12(t-1)^{\frac1t}<\frac12t^{\frac1t}\leq  \frac{e^{1/e}}2<1$)}, for fixed $t$ and sufficiently large $n$, we have that
\[
{\rm ex}(n,K_{t,t})\leq \,n^{2-\frac1t}.
\]
\end{theorem}
With this theorem in hand, we prove
\begin{theorem}
The hereditary class property $\Sigma=$``weakly sparse'' is a decomposition horizon.
\end{theorem}
\begin{proof}
Let $\mathscr C$ be a hereditary class with a  quasibounded-size weakly-sparse decomposition with parameter $2$. 
Then there exists a weakly sparse class $\mathscr D$
such that, for every $\epsilon>0$ and every large integer $n$, every graph $G\in\mathscr C$ with $n$ vertices has a vertex partition into at most~$n^\epsilon$ parts, each two parts inducing a subgraph in $\mathscr D$. 
(Indeed, from the definition there exists a decomposition into $\mathcal O(n^{\epsilon/2})$ parts, thus into at most $n^\epsilon$ parts for sufficiently large~$n$.)
As $\mathscr D$ is weakly sparse, there exists an integer $t$ such 
that $K_{t,t}$ is not a subgraph of a graph in $\mathscr D$.

Assume towards a contradiction that $\mathscr C$ is not weakly sparse.  
As $\mathscr C$ is hereditary, for every even integer $n$, $\mathscr C$ contains a graph $G_n$ of order $n$ with a $K_{n/2,n/2}$ spanning subgraph. 
Let $\epsilon<1/2t$ be a positive real.
Partitioning the vertex set of $G_n$ into $n^\epsilon$ classes, some two classes induce a subgraph $H_n\in\mathscr D$ with at least $\frac14n^{2-2\epsilon}$ edges. According to the K{\H o}v\'ari--S\'os--Tur\'an theorem, if  $n$ is sufficiently large, then~$H_n$ contains  a subgraph isomorphic to $K_{t,t}$ if it contains at least
$|H_n|^{2-\frac1t}\leq n^{2-\frac1t}$ edges. As $t<\frac{1}{2\epsilon}$,  for sufficiently large $n$ we have that $n^{2-\frac1t} < \frac14n^{2-2\epsilon}$, and so $H_n$ contains a subgraph isomorphic to $K_{t,t}$, contradicting our assumption on $\mathscr D$.
\end{proof}

For the next proposition, we use the following characterization of nowhere dense classes.

\begin{lemma}[See, e.g. \cite{Sparsity}]
\label{lem:ND}
For a hereditary class of graphs $\mathscr C$, the following are equivalent:
\begin{enumerate}
\item $\mathscr C$ is \emph{somewhere dense} (i.e.,  not nowhere dense);
\item for some integer $k$ and every integer $n$, the $k$th-subdivision of $K_n$ is a subgraph of some graph in $\mathscr C$; 
\item for some integer $k$ and $\epsilon>0$, for every integer $N$ there exists in $\mathscr C$ a graph that has a spanning subgraph which is a  $k$-subdivision of a graph $H$ with $n\geq N$ vertices and  at least $n^{1+\epsilon}$ edges.
\end{enumerate}
\end{lemma}

\begin{lemma}
	\label{lem:hor_and}
	Assume a (possibly non-hereditary) class $\mathscr C$ has quasibounded-size almost nowhere dense decompositions. Then $\mathscr C$ is almost nowhere dense.
\end{lemma}
\begin{proof}
Assume towards a contradiction that there exists an integer $k$ and a positive real $c>0$, such that for every integer $N$ there exist a graph $G\in\mathscr C$ and a graph $H$ with at least $N$ vertices and minimum degree at least $|G|^c$ such that some $\leq k$-th subdivision $H'$ of $H$ is a subgraph of $G$. 
Let $0<\epsilon<c/(2k+4)$, and let $p=k+2$. Let $\mathscr D_p$ be the almost nowhere dense class associated to the parameter~$p$.
We consider graphs $G$ and $H$ as above and a coloring  $\gamma:V(G)\rightarrow[|G|^\epsilon]$ of the vertices of $G$ such that every $p$ colors induce a subgraph in $\mathscr D_p$.  We color the edges of $H$ by the set of at most $k+2$ colors of the vertices in the paths of $H'$ corresponding to them. It follows that $H$ has an edge-monochromatic subgraph with average degree at least $|G|^c/|G|^{(k+2)\epsilon}=|G|^{c-(k+2)\epsilon}\geq |G|^{c/2}$. It follows that if 
$I$ is the set of $k+2$ colors and $G_I$ is the subgraph of $G$ induced by these colors, we have $\widetilde{\nabla}_{k+1}(G_I)\geq \|H\|/|H|\geq |G|^{c/2}\geq |G_I|^{c/2}$, contradicting $G_I\in\mathscr D_p$.
\end{proof}

\begin{corollary}
The hereditary class property ``nowhere dense'' is a decomposition horizon.
\end{corollary}
\begin{proof}
	This follows from Lemma~\ref{lem:hor_and} and the fact that a hereditary class of graphs is nowhere dense if and only if it is almost nowhere dense.
\end{proof}

We now prove the main results of this section, that ``dependent'' and ``stable'' are decomposition horizons. We will use the following characterization of dependence in hereditary classes.

\begin{theorem}[\!\!\cite{braunfeld2022existential}]
	\label{theorem:herNIP}
	For a hereditary class $\Cc$ the following conditions are equivalent:
	\begin{enumerate}
		\item $\Cc$ is dependent,
		\item $\Cc$ is monadically dependent,
		\item $\Cc$ does not admit pre-coding (by an existential formula).
	\end{enumerate}
\end{theorem}

We recall that a class $\Cc$ \emph{admits pre-coding} if there is a formula $\phi(\bar{x}, \bar{y}, z)$ (with parameters\footnote{\emph{Parameters} are additional free variables interpreted in a fixed way in each considered structure; intuitively, they behave like constants.}) and, for each $n \in \mathbb{N}$, a graph $G_n \in \Cc$ containing tuples $\bar{a}_i$ for $i \in [n]$, $\bar{b}_j$ for $j \in [n]$, and singletons $c_{i,j}$ for $(i,j) \in [n] \times [n]$ such that the following holds for every $(i,j) \in [n]\times[n]$.
\vspace{1mm}
\begin{itemize}
	\item $G_n \models \phi(\bar{a}_i, \bar{b}_j, c_{i,j})$
	\item $G_n \models \neg \phi(\bar{a}_i, \bar{b}_k, c_{i,j})$ for $k > j$
	\item $G_n \models \neg \phi(\bar{a}_h, \bar{b}_j, c_{i,j})$ for $h < i$
\end{itemize}
\vspace{1mm}
(The definition of pre-coding given in \cite{braunfeld2022existential} is slightly different, but easily seen to be equivalent.)

For a concrete example, one may consider the case where $G_n$ is a 2-subdivided complete bipartite graph with $n$ vertices on each side. Let $a_i$ be the $i^{th}$ vertex in the first part and~$b_j$ the $j^{th}$ vertex in the second. Let $c_{i,j}$ be the point adjacent to $a_i$ on the path from $a_i$ to~$b_j$. Then $\phi(x,y,z) := \exists w (xEz \wedge zEw \wedge wEy)$ witnesses pre-coding. If the proof of the next theorem is read with this case in mind, essentially no complexity is lost.

\begin{theorem}
	\label{theorem:NIPcover}
	The hereditary class property ``dependent'' is a decomposition horizon. 
\end{theorem}
\begin{proof}

	Let $\Cc$ be a hereditary class with quasibounded-size dependent decompositions.
	Then for every $p$ there is a hereditary dependent class $\mathscr D_p$ with the property that for every $\epsilon>0$ the vertex set of every graph $G\in\mathscr C$ can be partitioned into $N_p(G)=\mathcal O(|G|^{\epsilon})$ parts in such a way that the union of any $p$ classes induces a subgraph in $\mathscr D_p$.  Assume towards a contradiction that $\Cc$ is not dependent. Then, according to \cref{theorem:herNIP}, $\Cc$ admits pre-coding by an existential formula  $\phi(\bar{x},\bar{y},z)$. Let $p$ be the number of variables in $\phi$ (bound or free) plus the number of parameters. Consider an arbitrary integer $n$ and let $\epsilon>0$ be such that $\epsilon<1/(pn)$. By assumption, there exists, for every integer $m$,  a graph $G_m \in \Cc$ as in the definition of pre-coding, with parameters $\bar{d}$ in $G_m$.
	For $i,j\in [m]$, let $\bar{z}_{i,j}$ be the tuple of the vertices $\bar{a}_{i}\bar{b}_{j}c_{i,j}\bar{w}_{i,j}\bar{d}$, where $\bar{w}_{i,j}$ is a tuple of vertices witnessing the existential quantifiers in $G_m\models\phi(\bar{a}_{i},\bar{b}_{j},c_{i,j})$. Note that $|\bar{z}_{i,j}| = p$.
	Let $Z=\bigcup_{i,j\in [m]}(\bigcup \bar{z}_{i,j})$.
	As $\phi$ is existential, it follows that we may assume $G_m = G_m[Z]$, and thus that $|G_m|\leq pm^2$. (Note that we have also $\ |G_m|\geq m^2$ as all the vertices $c_{i,j}$ are distinct.)
	As $ N_p(G)=\mathcal O(|G|^{\epsilon/3})$,  for sufficiently large $m$ we have 
	$N_p(G_m)<m^\epsilon$.  Consider a partition of $V(G_m)$ into $N=N_p(G_m)$ parts $V_1,\dots,V_N$ such that the union of any $p$ parts belongs to $\Dd_p$. 
	Let $Z_k$ be the ``$k^{th}$-slice'' of~$Z$, i.e. $Z_k = \bigcup_{{i,j} \in [m]} \bar{z}_{i,j}(k)$. By the pigeonhole principle, for each $k \in [p]$ there is some~$V_{\ell_k}$ containing at least an $m^{1-\epsilon}$-proportion of $Z_k$. Taking these all together, there exists a subset $I\in\binom{[N]}{p}$ such that $G_m[\bigcup_{i \in I} V_i]$ includes $\bar{z}_{i,j}$ for at least $m^{2-p\epsilon} > m^{2-\frac{1}{n}}$ pairs $(i,j)\in [m]\times [m]$.  According to the
	K{\H o}v\'ari--S\'os--Tur\'an theorem (\Cref{thm:KST}), there exist (assuming~$m$ is sufficiently large) subsets $S,T\subseteq[m]$ of size at least~$n$ such that for all $(i,j)\in S\times T$, we have $\bar{z}_{i,j}\subseteq G_m[\bigcup_{i \in I} V_i]$. In particular, $\phi$ defines a pre-coding configuration of size $n$ in $G_m[\bigcup_{i \in I} V_i]$. As $n$ is arbitrary and as $G_m[\bigcup_{i \in I} V_i]\in\Dd_p$, we infer that $\phi$ witnesses that $\Dd_p$ admits pre-coding, contradicting the assumption that $\Dd_p$ is dependent.
\end{proof}

The special case of bounded-size bounded linear cliquewidth decompositions settles positively the conjecture that every class with low linear-rankwidth covers is monadically dependent.
Similarly, classes with low rank-width colorings (introduced in \cite{kwon17}) are monadically dependent.

We now move on to the analogue of \Cref{theorem:NIPcover} for stable decompositions. 
We shall need  the following result.

\begin{theorem}[\!\!\cite{nevsetril2020rankwidth}]
	\label{theorem:HG}
	A monadically dependent class of graphs is monadically stable if and only if it has a stable edge relation, that is, if it excludes some half-graph as a semi-induced subgraph. 
\end{theorem}

\begin{theorem}
	\label{theorem:stable_cover}
	Let $\mathscr C$ be a hereditary class of graphs. 
	Assume that $\Cc$ has quasibounded-size dependent decompositions, as well as a quasibounded-size stable decomposition for parameter~$2$. Then $\Cc$ is stable. 
	
	In particular, the hereditary class property ``stable'' is a decomposition horizon. 
\end{theorem}
\begin{proof}
	According to \cref{theorem:NIPcover} the class $\Cc$ is dependent. Being hereditary, it is monadically dependent by Theorem~\ref{thm:LB}.
	Assume towards a contradiction that~$\Cc$ is not stable. As $\Cc$ is monadically dependent and unstable, graphs in~$\Cc$  contain (according to \cref{theorem:HG}) arbitrarily large semi-induced half-graphs. As~$\Cc$ is hereditary, $\Cc$ contains, for each integer $m$ a graph $H_m$ on $2m$ vertices containing a half-graph of order~$m$ as a semi-induced subgraph. Consider any integer $n$. Then there exists an integer~$m$ such that $m/(f_2(|H_m|)^2)\geq n$ (where $f_2(|H_m|)$ is an upper bound on the number of parts in the decomposition of $H_m$ for parameter~$2$). 
	It follows (by a pigeonhole argument, picking one side of $H_m$ and coloring each point by the pair of its part in the decomposition and the part of the point at the same level on the other side) that $H_m$ contains a $2$-colored graph containing a half-graph of order $n$ as a semi-induced subgraph, which contradicts the assumption that $\Cc$ has a quasibounded-size stable decomposition with parameter $2$. 
\end{proof}

\begin{remark}
	We remark that \cref{theorem:NIPcover} actually holds for hereditary classes of relational structures, and the proof goes through without change. Similarly, \Cref{theorem:stable_cover} holds for hereditary classes of relational structures, with the parameter 2 replaced by the maximum arity $r$ of a relation in the language (if it exists). The proof requires a version of \cref{theorem:HG} for relational structures, which is provided in \cite{braunfeld2022existential}.
\end{remark}

\section{Decompositions of hereditary stable classes}
\label{sec:decst}

The aim of this section is to introduce all the techniques and results  needed to prove \Cref{thm:stSD} in Section~\ref{sec:main}, that is, that  a hereditary class of graphs  is stable if and only if it has quasibounded-size bounded shrubdepth decompositions.	

The first tool is the use of neighborhood covers, which allows us to restrict the study to graphs with bounded diameters. The second is the use of flipping, which is central to the characterization of monadically stable class by  the so-called Flipper game. This offers a way to decompose the graphs in a monadically stable class with a bounded depth recursion.  Such a decomposition is given by the so-called  quasi-bush representations, which are in spirit very close to the quasi-bounded size bounded shrubdepth decompositions we are aiming for.

\subsection{Neighborhood covers}
The notion of (sparse) neighborhood covers was used in \cite{grohe2014deciding} to design a quasi-linear time algorithm for first-order model checking on nowhere dense classes. The existence of sparse neighborhood covers on monotone classes is strongly related to bounding of weak coloring numbers \cite{GKRSS18}.
Let us recall the definitions.

The \emph{weak diameter} of a subset $X$ of vertices of a graph $G$ is the maximum distance in $G$ between vertices in $X$. The \emph{strong diameter} of $X$ is the diameter of $G[X]$ (and $\infty$ if $G[X]$ is not connected).
Let $G$ be a graph and $r$ be a positive integer. 
The \emph{(closed) $r$-neighborhood}~$N_G^r[v]$ of a vertex $v$ in $G$ is the set of all the vertices at distance at most $r$ from $v$ in $G$.  The \emph{$r$-ball} of $G$ centered at $v$, denoted  $B_r(G,v)$,  is the subgraph of $G$ induced by all the vertices at distance at most $r$ from $v$: $B_r(G,v)=G[N_G^r[v]]$. More generally, if~$X$ is a subset of vertices of $G$, we  define
$N_G^r[X]=\bigcup_{v\in X}N_G^r[v]$ and $B_r(G,X)=G[N_G^r[X]]$.
Note that $B_r(G,X)$ does not need to be connected in general.

A family $\mathscr K$ of subsets of $V(G)$  is a \emph{distance-$r$ neighborhood cover} of $G$ if, for every $v\in V(G)$, the $r$-neighborhood of $v$ in $G$ is included in some $C\in\mathscr K$. We define the \emph{weak (strong) diameter} of~$\mathscr K$ as the maximum weak (strong) diameter of the sets in $\mathscr K$, and the \emph{overlap} of $\mathscr K$ as the maximum number of sets in $\mathscr K$ a vertex of $G$ belongs to:
${\rm overlap}(\mathscr K):=\max_{u\in V(G)}\bigl|\{C\in\mathscr K\colon u\in C\}\bigr|$.
Following~\cite{dreier2023firstorder}, for a graph $G$ we call a subset $X$ of  $V(G)$ \emph{compact} if there exists a vertex $u$ of $G$ with $X\subseteq N_G[u]$. A partition $\mathscr Z$ of the vertex set of $G$ is called compact if every element of $\mathscr Z$ is compact. A distance-$1$ neighborhood cover $\mathscr K$ is \emph{compact} 
if there exists a compact partition $\mathscr Z$ of $V(G)$ with
$\mathscr K=\{N_G[Z]\colon Z\in\mathscr Z\}$.
The \emph{incidence graph}  of a neighborhood cover $\mathscr K$ of a graph $G$ is the bipartite graph ${\rm Inc}(\mathscr K)=(\mathscr K, V(G), F)$, where $C\in\mathscr K$ is adjacent to $v\in V(G)$, that is, $\{C,v\}\in F$, if $v\in C$.
%
%
Note that if $r\leq r'$, every distance-$r'$ neighborhood cover of $G$ is also a distance-$r$ neighborhood cover of $G$. 

The existence of sparse neighborhood covers with small overlap can be used to characterize those  monotone classes of graphs that have bounded expansion or are nowhere dense.

\begin{theorem}[\cite{Japan04}]
	A monotone class of graphs has bounded expansion if and only if for every radius $r$ there exists integers $c$ and $N$ such that  every graph $G\in\mathscr C$ has a distance-$r$ neighborhood cover with diameter at most $cr$ and overlap at most $N$.
\end{theorem}

\begin{theorem}[\cite{GKRSS18}]
	A monotone class of graphs is nowhere dense if and only if for every radius $r$ there exists an integer $c$ such that  every $n$-vertex graph $G\in\mathscr C$ has a distance-$r$ neighborhood cover with diameter at most $cr$ and overlap $\mathcal O_{\mathscr C,r,\epsilon}(n^\epsilon)$, for every $\epsilon>0$.
\end{theorem}

The existence of sparse neighborhood covers has further been extended to structurally nowhere dense classes \cite{Dreier2023} and eventually to monadically stable classes.
\begin{theorem}[\cite{dreier2023firstorder}]
	\label{thm:nc}
	Let $\mathscr C$ be a monadically stable class of graphs. Then for every $r\in\mathbb N$ and $n$-vertex graph $G\in\mathscr C$ there exists a distance-$r$ neighborhood cover of $G$ with weak diameter at most $4r$ and whose overlap is in $\mathcal O_{\mathscr C,r,\epsilon}(n^\epsilon)$ for every $\epsilon>0$. Moreover, there is an algorithm that computes such a neighborhood cover in time $\mathcal O_{\mathscr C,r,\epsilon}(n^{4+\epsilon})$, for every $\epsilon>0$.
\end{theorem}

As noted in \cite[Appendix C.2]{dreier2023firstorder}, the distance-$1$ neighborhood
cover of $G$ constructed in the proof of Theorem~\ref{thm:nc} is compact.
Hence, as the distance-$r$ neighborhood cover is obtained by constructing a distance-$1$ neighborhood cover in $G^r$, we have (as $N_{G^r}[X]=N_G^r[X]$ for every subset $X$ of vertices):
\begin{theorem}[\cite{dreier2023firstorder}]
		\label{thm:ncbis}
	Let $\mathscr C$ be a monadically stable class of graphs. Then for every $r\in\mathbb N$ and $n$-vertex graph $G\in\mathscr C$  there exists a partition $\mathscr I$ of $V(G)$ and a mapping $\rho:\mathscr I\rightarrow V(G)$ such that for every $I \in\mathscr I$ we have $I\subseteq N_G^r[\rho(I)]$ and 
\[
	\max_{v\in V(G)}|\{I\in\mathscr I\colon v\in N_G^r[I]\}|\in \mathcal O_{\mathscr C,r,\epsilon}(n^\epsilon)\qquad(\forall\epsilon>0.)
\]
 Moreover, there is an algorithm that computes such a partition in time $\mathcal O_{\mathscr C,r,\epsilon}(n^{4+\epsilon})$, for every $\epsilon>0$.
\end{theorem}

\begin{remark}
	\label{rem:nc}
It follows from $I\subseteq N_G^r[\rho(I)]$ that $\rho(I)\in N_G^r(I)$. Hence, $\{G[N_G^r[I]]\colon I\in \mathscr I\}$ is a distance-$r$ neighborhood cover with \emph{strong} diameter at most $4r$ and overlap 
$\mathcal O_{\mathscr C,r,\epsilon}(|G|^\epsilon)$, for every $\epsilon>0$.
\end{remark}

We consider the following strengthening of distance-$r$ neighborhood covers.

\begin{definition}
An \emph{induced distance-$r$ neighborhood cover} of a graph $G$ 
is a family $\mathscr K $ of 
induced subgraphs of $G$, called \emph{clusters}, such that
	for every vertex $u$ of $G$, there is a cluster $C \in\mathscr  K$ that contains the $r$-neighborhood of $u$.
The \emph{weak} (\emph{strong}) \emph{diameter} of $\mathscr K$ is the maximum weak (strong) diameter of the connected components of the clusters of 
$\mathscr K$. The \emph{size} of $\mathscr K$ is the cardinality of $\mathscr K$.
\end{definition}

Note that the connected components of any induced distance-$r$ neighborhood cover $\mathscr K$ 
of weak diameter $d$ and size $k$ yield 
a distance-$r$ neighborhood cover with diameter $d$ and overlap $k$.

Let us now remark that the existence of induced distance-$r$ neighborhood covers can be deduced from the existence of special vertex colorings for class properties that are preserved by transductions.
For an integer $\ell$, a \emph{weak diameter-$\ell$ coloring} of a graph $G$ is a vertex coloring of $G$ such that each monochromatic connected subgraph has weak diameter at most~$\ell$~\cite{DVORAK2023103845}. 

\begin{lemma}
	\label{lem:inc_from_col}
	Let $\Pi$ be a class property preserved by transduction. 
	Then the following properties are equivalent:
	\begin{enumerate}[(i)]
		\item for every integer $r$ and every class $\mathscr C\in\Pi$, every graph $G\in\mathscr C$ has  induced distance-$r$ neighborhood covers with weak diameter $\mathcal O_{\mathscr C}(r)$ and size  $\mathcal O_{\mathscr C,r,\epsilon}(1)$ (resp.\ $\mathcal O_{\mathscr C,r,\epsilon}(|G|^\epsilon)$ for every $\epsilon>0$);
		\item for every class $\mathscr C\in\Pi$ there exists an integer $d$ such that every graph $G\in\mathscr C$ has a weak diameter-$d$  coloring by $\mathcal O_{\mathscr C}(1)$  colors (resp.\ with $\mathcal O_{\mathscr C,\epsilon}(|G|^\epsilon)$ colors for every $\epsilon>0$ ).
	\end{enumerate}
\end{lemma}
\begin{proof}
	(i) $\Rightarrow$ (ii): Consider $r=1$ and let $C_1,\dots,C_k$ be the clusters of $G$.
	Color a vertex $v$ by the minimum integer $i$ such that  $C_i$ includes the (closed) neighborhood of $v$. Then, each monochromatic connected graph has weak diameter bounded by some constant $d$.
	
	(ii) $\Rightarrow$ (i): Let $\mathscr D=\{G^{2r+1}\colon G\in\mathscr C\}$. As $\mathscr D$ is a transduction of $\mathscr C$, we have $\mathscr D\in\Pi$. Thus, there exists an  integer $d$ such that every graph in $\mathscr D$ has a coloring by $f_{\mathscr D}(|G|)$ colors without monochromatic connected components of diameter greater than $d$, where either $f_{\mathscr D}(x)\in\mathcal O(1)$ or $f_{\mathscr D}(x)\in\mathcal O(x^\epsilon)$ for every $\epsilon>0$.
	For $G\in\mathscr C$, we transfer this coloring from~$G^{2r+1}$ to  $G$.
	Let $C_i$ be the subgraph of $G$ induced by the union of all the $B_r(v)$, where~$v$ has color $i$.  
	Consider an induced path $P=(u_1,\dots,u_n)$, whose vertices belong to~$C_i$.  For each $j\in [n]$, there exists a vertex $v_j$ with color $i$ such that ${\rm dist}(u_j,v_j)\leq r$. Hence, 
	${\rm dist}(v_j,v_{j+1})\leq 2r+1$, meaning that $v_j$ and $v_{j+1}$ are either equal or adjacent in~$G^{2r+1}$. It follows that all the~$v_j$ belong to a same monochromatic component of $G^{2r+1}$, which has weak diameter at most~$d$. Thus, there exists a vertex $v$ with color $i$ such that ${\rm dist}(v,v_j)\leq (2r+1)d$. Hence, ${\rm dist}(u_1,u_n)\leq 2(r+(2r+1)d)\leq (3d+2)r$. 
\end{proof}

As an example of application of Lemma~\ref{lem:inc_from_col} outside the realm of stable classes of graphs, let us mention the following.

\begin{corollary}
	\label{cor:lrw}
	Every class $\mathscr C$ with bounded linear rankwidth has induced distance-$r$ neighborhood covers with weak diameter $12r+4$ and size $\mathcal O_{\mathscr C}(1)$.
\end{corollary}
\begin{proof}
	The property of having bounded linear rankwidth is preserved by transduction (as it is equivalent to being a transduction of a linear order). Moreover, every graph with linear rankwidth $t$ has an $f(t)$-vertex coloring such that every color class induces a cograph (hence has diameter at most $2$)~\cite{msrw}. The bound on the diameter now comes from the proof of 
	Lemma~\ref{lem:inc_from_col}.
\end{proof}

We prove the following strengthening of Theorem~\ref{thm:nc}, which we shall use later in the case $r=1$.

\begin{figure}[ht]
\begin{center}
	\includegraphics[width=.7\textwidth]{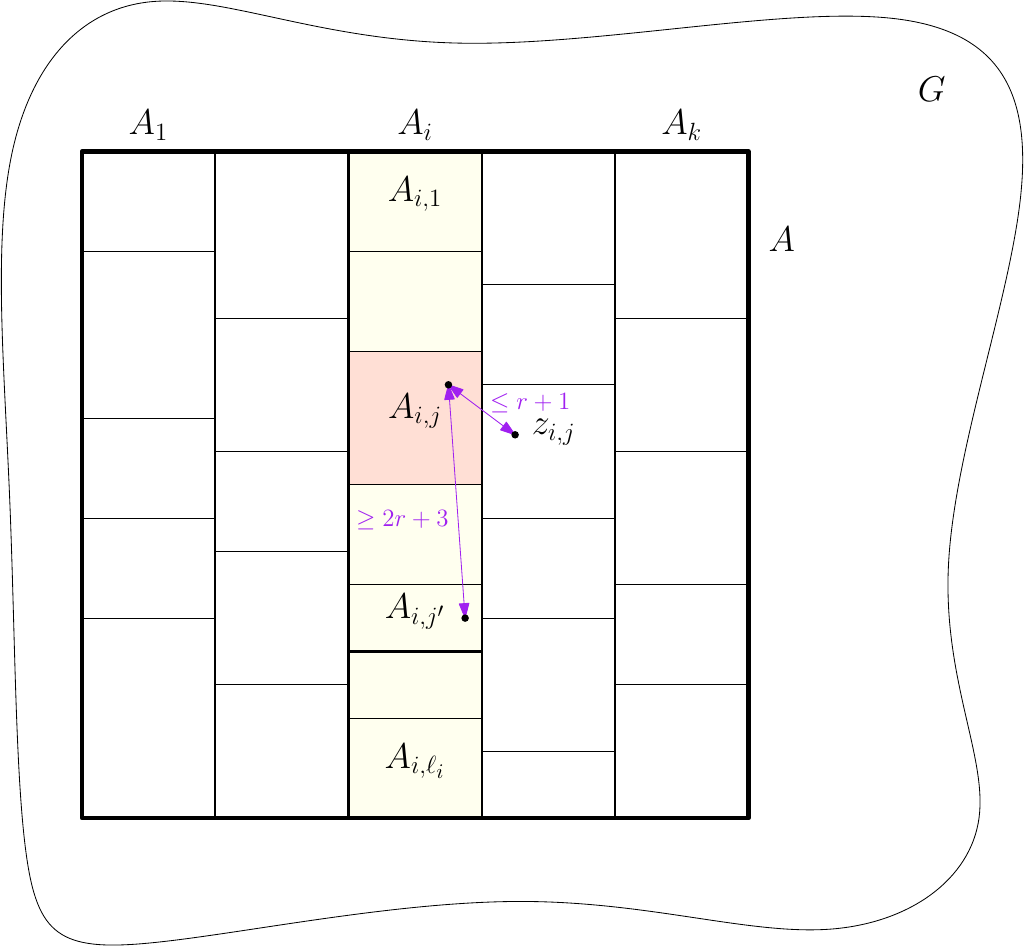}
	\caption{A partition of a subset $A$ of vertices of a graph $G$ with parameter $r$, as obtained by \Cref{thm:incg}. 
		To each part $A_{i,j}$ is associated a vertex $z_{i,j}$ (possibly not in $A$). The distance between $z_{i,j}$ and any vertex in $A_{i,j}$ is at most $r+1$, while the distance between any vertex in $A_{i,j}$ and any vertex in $A_{i,j'}$ (for $j'\neq j$) is at least $2r+3$.
	Consequently, $A_{i,j}=N_G^{r+1}[z_{i,j}]\cap A_i$.
	}
	\label{fig:cluster}
\end{center}
\end{figure}

\begin{theorem}
	\label{thm:incg}
	Let $\mathscr C$ be a monadically stable class of graphs and let $r\in\mathbb N$. Then for every $\epsilon>0$,  every $n$-vertex graph $G\in\mathscr C$, and every subset $A$ of vertices of $G$, there exists a partition $A_1,\dots,A_k$ of $A$ into $k\in\mathcal O_{\mathscr C,r,\epsilon}(n^\epsilon)$ parts and, for each $1\leq i\leq k$ a partition $A_{i,1},\dots,A_{i,\ell_i}$ of $A_i$
	such that (See Fig.~\ref{fig:cluster}):
\begin{enumerate}
	\item for every $1\leq i\leq k$ and every $1\leq j_1<j_2\leq \ell_i$, the minimum distance in $G$ between a vertex in $A_{i,j_1}$ and a vertex in $A_{i,j_2}$ is at least $2r+3$;
	\item 	for every $1\leq i\leq k$ and every $1\leq j\leq \ell_i$ there exists a vertex $z_{i,j}$ of $G$ with 
	$A_{i,j}\subseteq N_G^{r+1}[z_{i,j}]$. 
\end{enumerate}
	
	Moreover, there is an algorithm that computes such a partition and the vertices $z_{i,j}$ in time $\mathcal O_{\mathscr C,r,\epsilon}(n^{4+\epsilon})$, for every $\epsilon>0$.
\end{theorem}

\begin{remark}
	\label{rem:weak_diam}
Note that $\mathscr K=\{N_G^r[A_{i}]\colon 1\leq i\leq N(G,r+1)\}$ is an induced distance-$r$ neighborhood cover with size $N(G,r+1)\in\mathcal O_{\mathscr C,r,\epsilon}(|G|^\epsilon)$ for every $\epsilon>0$. We shall call the graphs $G[N_G^r[A_i]]$ the \emph{clusters} of the induced neighborhood cover, and the graphs
$G[N_G^r[A_{i,j}]]$ the \emph{components} of the cluster $G[N_G^r[A_i]]$. 
(Note that $G[N_G^r[A_i]]$ is the disjoint union of the graphs $G[N_G^r[A_{i,j}]]$.)
We further call
$z_{i,j}$ the \emph{handle} of the component $G[N_G^r[A_{i,j}]]$.
Notice that $G[N_G^r[A_{i,j}]\cup\{z_{i,j}\}]$ is a connected induced subgraph of $G$ with diameter at most $4r+2$.
It follows that the induced distance-$r$ neighborhood cover  $\mathscr K$ has weak diameter at most $4r+2$.
\end{remark}

%

In order to prove Theorem~\ref{thm:incg} we need two preliminary results. The first result concerns the incidence graph of the distance-$r$ neighborhood cover constructed in the proof of Theorem~\ref{thm:ncbis}.

\begin{theorem}[\cite{dreier2023firstorder}]
	\label{thm:nc_inc}
	Let $\mathscr C$ be a monadically stable class of graphs and $r\in\mathbb{N}$ be fixed.
	For every graph $G\in\mathscr C$, let $\mathscr K_{G,r}$ be the distance-$r$ neighborhood cover of $G$ given by  Theorem~\ref{thm:ncbis} (as indicated in \Cref{rem:nc}).  Then
	the class $\{{\rm Inc}(\mathscr K_{G,r})\colon G\in\mathscr C\}$  is almost nowhere dense.
\end{theorem}

The second result is a technical lemma bounding the degeneracy of the square of a bipartite graph.
Recall that the degeneracy $\rm deg(G)$ of a graph $G$ is the smallest integer $d$ such that all subgraphs $H\subseteq G$ have a vertex of degree (in $H$) at most $d$.

\begin{lemma}
	\label{lem:degsq}
	Let $G=(A,B,E)$ be a bipartite graph, let $\Delta_B$ be the maximum degree of the vertices in $B$, and let $G^2[A]$ be the subgraph of $G^2$ induced by $A$. Then the degeneracy ${\rm deg}(G^2[A])$ of $G^2[A]$ is bounded by
	\[
	{\rm deg}(G^2[A])\leq 2(\Delta_B-1)\nabla_1(G).
	\]
\end{lemma}
\begin{proof}
	Consider a coloring $c$ of the edges of $G$ with $\Delta_B$ colors, such that no two edges incident to the same vertex in $B$ have the same color. For $1\leq i<\Delta_B$, let $H_i$ be the subgraph of $G^2[A]$ obtained by contracting all the edges colored $i$, removing parallel edges, and deleting the vertices in $B$. Note that the edges colored $i$ induce a star forest in $G$, hence $H_i$ is a depth-$1$ minor of $G$. Moreover, every edge of $G^2[A]$ is covered by at least one of the $H_i$. Hence, for every non-empty subset $X$ of $A$, we have $|E(G^2[X])|\leq \sum_i |E(H_i[X])|\leq (\Delta_B-1)\,\nabla_1(G)\,|X|$. Hence,
	$G^2[A]$ is $(2(\Delta_B-1)\nabla_1(G))$-degenerate.
\end{proof}


\begin{proof}[Proof of Theorem~\ref{thm:incg}]
	Let $\mathscr I_{G,r+1}$ be the partition of the graph $G\in\mathscr C$ given by Theorem~\ref{thm:ncbis}, and let $\mathscr K_{G,r+1}=\{N_G^{r+1}[I]\colon I\in\mathscr I\}$ be the derived distance-$(r+1)$ neighborhood cover (See \Cref{rem:nc}).
	
	Recall that $\{{\rm Inc}(\mathscr K_{G,r+1})$ is the bipartite graph with parts $\mathscr K_{G,r+1}$ and $V(G)$, where $C\in \mathscr K_{G,r+1}$ is adjacent to $v\in V(G)$ if $v\in C$. According to Theorem~\ref{thm:ncbis}, every vertex of $G$ belongs to $\mathcal O_{\mathscr C,r,\epsilon}(|G|^\epsilon)$ sets in  ${\rm Inc}(\mathscr K_{G,r+1})$. Moreover, according to Theorem~\ref{thm:nc_inc}, the class 
	$\{{\rm Inc}(\mathscr K_{G,r+1})\colon G\in\mathscr C\}$  is almost nowhere dense, hence $\nabla_1({\rm Inc}(\mathscr K_{G,r+1}))\in\mathcal O_{\mathscr C,r,\epsilon}(|G|^\epsilon)$.
	
	Thus, according to Lemma~\ref{lem:degsq}, the graphs
	${\rm Inc}(\mathscr K_{G,r+1})^2[\mathscr K_{G,r+1}]$ are $\mathcal O_{\mathscr C,r,\epsilon}(|V(G)|^{2\epsilon})$-degenerate. Note that two sets $C_1, C_2\in\mathscr K_{G,r+1}$ are adjacent in  ${\rm Inc}(\mathscr K_{G,r+1})^2[\mathscr K_{G,r+1}]$  if and only if 
	$C_1\cap C_2\neq\emptyset$. Hence, a greedy coloring of $\mathscr K_{G,r+1}$ colors the sets in $\mathscr K_{G,r+1}$ with  $k \in \mathcal O_{\mathscr C,r,\epsilon}(|V(G)|^{2\epsilon})$ colors in such a way that any two sets with the same color are disjoint. Let $c$ denote this coloring. We transfer this coloring to $\mathscr I$ by defining $\gamma(I)=c(N_G^{r+1}[I])$.
	Hence, for all distinct $I_1,I_2\in \mathscr I$ with $\gamma(I_1)=\gamma(I_2)$ we have $N_G^{r+1}(I_1)\cap N_G^{r+1}(I_2)=\emptyset$, thus the minimum distance in $G$ between a vertex in $I_1$ and a vertex in $I_2$ is at least $2r+3$.
	Moreover, by construction, for every $I\in\mathscr I$, there exists a vertex $\rho(I)$ such that $I\subseteq N_G^{r+1}[\rho(I)]$.
	For sake of simplicity, let us denote by $I_{i,j}$ the $j$th set in $\mathscr I$ with color $i$.
	For $1\leq i\leq k$, we define 
	$A_{i,j}=A\cap I_{i,j}$ and $A_i=\bigcup_j A_{i,j}$. 
	Then, the conditions of Theorem~\ref{thm:incg} are satisfied.
	
%
	
	The computation of ${\rm Inc}(\mathscr K_{G,r+1})^2$, its greedy coloring, and the construction of the induced neighborhood cover does not increase the time complexity. 
\end{proof}

%
%
%

\subsection{SC-depth and Flipper game}
\label{sec:Flipper}
Let $G$ be a graph and let $F\subseteq V(G)$.
We denote by ${G}\oplus F$ the graph $G'$ obtained by complementing the edges with both endvertices in~$F$, i.e., $G'$ has vertex set~$V(G)$ where
$x\neq y$ are adjacent in $G'$ if either
(i) $\{x,y\}\in E(G)$ and $\{x,y\}\not\subseteq F$, or (ii)~$\{x,y\}\not\in E(G)$ and $\{x,y\}\subseteq F$.

The {\em SC-depth} of a graph $G$, denoted $\sd(G)$, is defined inductively in a very similar way to treedepth, by replacing vertex deletion by subset complementation \cite{ganian2012trees}:
\begin{itemize}
	\item $\sd(K_1)=1$;
	\item $\sd(G+H)=\max(\sd(G),\sd(H))$;
	\item 	$\sd(G)=1+\min_{F\subseteq V(G)}\sd(G\oplus F)$ if $G$ is connected and $G\not\simeq K_1$.
\end{itemize}

A notable property of SC-depth is that a class has bounded shrubdepth if and only if it has bounded SC-depth \cite{ganian2012trees}.

The \emph{Flipper game} is a two-player game played on a graph parametrized by an integer $r$, the \emph{radius} of the game. 
The two players are Flipper and Keeper. They play alternately.  In our version of the game, Keeper plays first. Keeper chooses a vertex $v$ and replaces the current graph by the ball of radius~$r$ around~$v$. 
Then Flipper  chooses a subset of vertices $F$ and \emph{flips} it, that is, it replaces the current graph~$G$ by $G\oplus F$. 
Flipper wins when the graph is reduced to $K_1$. The aim of Keeper is to delay Flipper's win as much as possible.
The Flipper game can also be seen as a variant of the Splitter game introduced in \cite{grohe2014deciding} to devise a model checking algorithm for nowhere dense classes, where subset flips replace vertex deletions.
This similarity justifies the conjecture that this game characterizes those hereditary classes of graphs that are stable, which was recently proved. We need only one direction.

\begin{theorem}[\cite{gajarsky2023flipper}]
	\label{thm:flipper}
	Let $\mathscr C$ be a hereditary,  stable class of graphs.
	Then, for every integer $r\geq 1$, there is a constant ${\rm FG}_{\mathscr C}(r)$ such that Flipper wins the flipper game of radius $r$ on graphs in $\mathscr C$ in at most ${\rm FG}_\mathscr C(r)$ rounds.
\end{theorem}

We  redesign the Flipper game as a local version of the SC-depth. 

\begin{definition}
	\label{def:Frank}
 For an integer $r$, let  $\Frank_r(G)$ be inductively defined by
\begin{itemize}
	\item $\Frank_r(K_1)=1$,
	\item $\Frank_r(G)=\max_{v\in V(G)}\Frank_r(B_r(G,v))$,
	\item $\Frank_r(G)=1+\min_{F\subseteq V(G)}\Frank_r(G\oplus F)$ if $G$ has radius at most $r$ and $G\not\simeq K_1$.
\end{itemize}
\end{definition}

Note that this graph invariant basically mimics the Flipper game, and that  $\sd(G)=\Frank_\infty(G)$.
We now recall a monotonicity property of the Flipper game, stated here in terms of $\Frank_r$.
\begin{lemma}
	\label{lem:fmon}
	If $H\subseteq_i G$, then $\Frank_r(H)\leq \Frank_r(G)$.
\end{lemma}
\begin{proof}
	We prove the statement by induction on $\Frank_r(G)$. The statement obviously holds if $\Frank_r(G)=1$. Otherwise, let $\Frank_r(G)=i>1$.
	As $H\subseteq_i G$, for every $v\in V(H)$ we have $B_r(H,v)\subseteq_i B_r(G,v)$. 
	Hence, we can restrict our analysis to the case where both $G$ and  $H$ have radius at most $r$, as the general case will follow by $\Frank_r(H)=\max_{v\in V(G)}\Frank_r(B_r(H,v))\leq \max_{v\in V(G)}\Frank_r(B_r(H,v))=\Frank_r(G)$.
	If both $G$ and~$H$ have radius at most $r$, then 
	there exists a subset $F$ of $V(G)$ such that $\Frank_r(G)=1+\Frank_r(G\oplus F)$.
	As $H\oplus (F\cap V(H))\subseteq_i G\oplus F$, the induction hypothesis implies that $\Frank_r(H)\leq 1+\Frank_r(H\oplus (F\cap V(H)))\leq 1+\Frank_r(G\oplus F)=\Frank_r(G)$.
\end{proof}
With these definitions, Theorem~\ref{thm:flipper} can be restated as 
\begin{theorem}
	Every hereditary stable class has bounded $\Frank_r$ for every integer $r$.
\end{theorem}

\subsection{Quasi-bush representations}

The computation of quasibounded-size bounded shrubdepth decompositions for monadically stable classes of graphs will be based on quasi-bush representations, which were introduced in~\cite{dreier2022treelike}.
A \emph{quasi-bush} is a binary structure consisting of a bounded height rooted tree $Y$, a set~$D$ of directed edges (called \emph{pointers}) from the leaves of $Y$ to internal nodes
of $Y$,  a labeling function $\lambda$, from the leaves of $Y$ to a finite set $\Lambda$  and a  labeling function $\lambda^D$ from $D$ to the power set of $\Lambda$.
A quasi-bush $B$ encodes a graph $G(B)$ whose vertex set is the set of leaves of $Y$ as follows:
a vertex $u$ is adjacent to a vertex $v$ if there exists a pointer $(u,z)$ from $u$ to an ancestor $z$ of $v$ in the tree, such that 
\begin{itemize}
	\item $z$ is the lowest ancestor  of $v$ such that 
	$(u,z)$ is a pointer;
	\item the label of $v$ belongs to the label of $(u,z)$, that is: $\lambda(v)\in\lambda^D(u,z)$.
\end{itemize}

In this paper, instead of giving a label $\lambda(v)$ to a vertex $v$ and a label $\lambda^D(e)$ to an  arc $e$, we associate to both a binary vector ($\mathbf b(v)$ and $\mathbf b(e)$, respectively) and replace the condition $\lambda(v)\in\lambda^D(e)$ by the condition $\langle\mathbf b(v),\mathbf b(e)\rangle=1$.

In \cite{dreier2023firstorder} it was proved that every structurally nowhere dense class admits almost nowhere dense quasi-bush representations, and the following conjecture was proposed.
\begin{conjecture}[{\cite[Conjecture 4]{dreier2023firstorder}}]
For every monadically stable class of graphs $\mathscr C$, there is an almost nowhere dense class $\mathscr B$ of
quasi-bushes that encode graphs from $\mathscr C$.
\end{conjecture}

In this section, we prove this conjecture.
The construction of a  quasi-bush representation will be based on two ingredients: induced neighborhood covers and the $\Frank_r$ invariants. It will consist of two steps: the construction of a decomposition tree, and the definition of the pointers.

We now define a decomposition tree for graphs in a hereditary stable class $\mathscr C$, which is based on Theorems~\ref{thm:nc_inc} and~\ref{thm:flipper}.
The nodes of the decomposition tree will be indexed using words $\alpha,\beta\in\mathbb{N}^{<\omega}$. 
We denote the empty word by $\lambda$, in order to avoid confusion with the real value $\epsilon$ appearing in the definition of a quasibounded-size $\Pi$ decomposition.

Each node of the tree is denoted using two words of the same length as indices.
The common length of the indices is the \emph{level} of the node in the tree.
Consequently, the $i$th level of the tree will be the set of all the nodes of the tree at level $i$.
The root of the tree is $\nu_\lambda^\lambda$, and the parent of $\nu_{\alpha i}^{\beta j}$ is 
$\nu_{\alpha}^{\beta}$. To each node $\nu_{\alpha}^{\beta}$ is associated a pair $(G_{\alpha}^{\beta} , A_\alpha^{\beta})$, where $G_\alpha^{\beta}$ is a graph and $A_\alpha^{\beta}$ is a non-empty subset of vertices of $G_\alpha^{\beta}$.
The pair associated to the root node is $(G_\lambda^\lambda, A_\lambda^\lambda)$, where $G_\lambda^\lambda=G$ and $A_\lambda^\lambda=V(G)$.

If the graph $G_\alpha^{\beta}$ associated to a  node~$\nu_{\alpha}^{\beta}$ of the decomposition tree contains a single vertex, then $\nu_{\alpha}^{\beta}$ is a leaf of the tree. (Note that if $\nu_{\alpha}^{\beta}$ is a leaf of the tree, then $A_\alpha^\beta=V(G_\alpha^\beta)$.)

For each node~$\nu_{\alpha}^{\beta}$ of the decomposition tree, whose associated graph $G_\alpha^{\beta}$ contains more than one vertex, 
we consider a partition $A_{\alpha 1}^{\beta},\dots,A_{\alpha k}^{\beta}$ of $A_\alpha^{\beta}$,  with each $A_{\alpha i}^\beta$  itself partitioned into $A_{\alpha i}^{\beta 1},\dots, A_{\alpha_i}^{\beta n_i}$, as obtained by Theorem~\ref{thm:incg} (with $r=1$). 
Following \Cref{rem:weak_diam}, we define 
the  components 
$C_{\alpha i}^{\beta j}=G_\alpha^{\beta}[N_{G_\alpha^{\beta}}[A_{\alpha i}^{\beta j}]]$ (for index~$j$ ranging from $1$ to  $n_i$)
of the cluster $G_\alpha^{\beta}[N_{G_\alpha^{\beta}}[A_{\alpha i}^{\beta}]]$ and  
 the handle $z_{\alpha i}^\beta$ of the component $C_{\alpha i}^\beta$. Let $\widehat C_{\alpha i}^\beta=G_\alpha^\beta[N^{3}_{G_\alpha^\beta}[z_{\alpha i}^\beta]]$. 
Note that~$C_{\alpha i}^\beta$ is an induced subgraph of $\widehat C_{\alpha i}^\beta$ and that $\widehat C_{\alpha i}^\beta$ has diameter at most $6$ (See \Cref{rem:weak_diam} again).
 By Definition~\ref{def:Frank}, there exists a subset~$\widehat{F}_{\alpha i}^{\beta j}$ of $V(\widehat{C}_{\alpha i}^{\beta j})$ such that
 $\Frank_6(\widehat{C}_{\alpha i}^{\beta j}\oplus \widehat{F}_{\alpha i}^{\beta j})<\Frank_6(\widehat{C}_{\alpha i}^{\beta j})$.
 Let $F_{\alpha i}^{\beta j}=\widehat{F}_{\alpha i}^{\beta j}\cap V({C}_{\alpha i}^{\beta j})$ and
  $G_{\alpha i}^{\beta j}={C}_{\alpha i}^{\beta j}\oplus F_{\alpha i}^{\beta j}$. Note that $G_{\alpha i}^{\beta j}$ is a induced subgraph of $\widehat{C}_{\alpha i}^{\beta j}\oplus \widehat{F}_{\alpha i}^{\beta j}$
  (with the same vertex set as $C_{\alpha i}^{\beta j}$, that is 
  $N_{G_\alpha^\beta}[A_{\alpha i}^{\beta j}]$).
According to Lemma~\ref{lem:fmon},
\[
	\Frank_6(G_{\alpha i}^{\beta j})\leq \Frank_6(\widehat{C}_{\alpha i}^{\beta j}\oplus \widehat{F}_{\alpha i}^{\beta j})<\Frank_6(\widehat{C}_{\alpha i}^{\beta j})\leq \Frank_6(G_\alpha^{\beta}).
\]

%

For each $i\in [k]$ and $j\in [n_i]$, we add to $\nu_{\alpha}^{\beta}$ a child $\nu_{\alpha i}^{\beta j}$. 
By construction, 
the sets $A_{\alpha i}^{\beta j}$ form a partition of $A_\alpha^{\beta}$. Moreover, 
$V(G_\alpha^{\beta})\supseteq \bigcup_{i,j} V(G_{\alpha i}^{\beta j})$.
Also note that the 
height of the tree is at most $h:=\max\{\Frank_6(G)\colon G\in\mathscr C\}$. 

\begin{fact}
	For every integer $0\leq \ell\leq h$, the sets $A_{\alpha,\beta}$ associated
	to leaves at level smaller than $\ell$ and nodes at level $\ell$ form a partition of $V(G)$.
	
	Consequently, for each $v\in V(G)$ there  is a unique leaf  $\rho(v)=\nu_\alpha^\beta$ with $A_\alpha^\beta=\{v\}$, and this mapping $\rho$ is a bijection from $V(G)$ to the set of all the leaves of the tree decomposition.
	\end{fact}
\begin{proof}
	We prove the statement by induction on $\ell$. The base case $\ell=0$ is trivial as $A_\lambda^\lambda=V(G)$. Assume that the statement has been proved for some $0\leq \ell<h$. By construction, the sets~$A_{\alpha i}^{\beta}$ with $|\alpha|=|\beta|=\ell$ form a partition of $A_\alpha^\beta$ and the sets $A_{\alpha i}^{\beta j}$ form a partition of $A_{\alpha i}^{\beta}$. It follows that the statement holds for $\ell+1$.
	
	The case where $\ell=h$ shows that the sets  $A_{\alpha}^\beta$ for leaves $\nu_{\alpha}^\beta$ form a partition of $V(G)$ into singletons, hence proving that the vertices of $G$ are in bijection with the leaves of the tree, as stated. 
\end{proof}

We define a coloring $\coloring$ on $V(G)$ by $\coloring(v)=\alpha$ if the leaf of the tree associated to $v$ has the form $\nu_{\alpha}^{\beta}$.

%
\begin{fact}
	\label{fact:disj}
	For every fixed $\alpha$ and every $\beta\neq \gamma$, the graphs $G_\alpha^{\beta}$ and $G_\alpha^{\gamma}$ are vertex-disjoint. Consequently, 	for every vertex $u$ and every word $\alpha$, there exists at most one graph~$G_\alpha^{\beta}$ with $u\in V(G_\alpha^{\beta})$.
\end{fact}
\begin{proof}
	Assume towards a contradiction that this property is not true, and consider a counter-example with $|\alpha|$ minimum. Let $v$ be a common vertex of $G_\alpha^{\beta}$ and $G_\alpha^{\gamma}$.
	As the graphs associated to the parents of $\nu_\alpha^{\beta}$ and $\nu_\alpha^{\gamma}$ do not form a counter-example (by the minimality of~$|\alpha|$), it follows that  $\nu_\alpha^{\beta}$ and $\nu_\alpha^{\gamma}$  are siblings with parent $\nu_\zeta^{\xi}$. However, this would imply that~$C_\alpha^{\beta}$ and $C_\alpha^{\gamma}$ are distinct  components of $G_\zeta^{\xi}[N_{G_\zeta^{\xi}}[A_{\alpha}^{\xi}]]$,   contradicting the hypothesis that they share a vertex.
\end{proof}

\begin{fact}
	\label{fact:distA}
	For every fixed $\alpha$ and every $\beta\neq \gamma$, the vertices in $A_\alpha^{\beta}$ and $A_\alpha^{\gamma}$ are at distance at least $5$.
\end{fact}
\begin{proof}
	Assume towards a contradiction that this property is not true, and consider a counter-example with $|\alpha|$ minimum. As the sets associated to the parents of $\nu_\alpha^{\beta}$ and $\nu_\alpha^{\gamma}$ do not form a counter-example (by the minimality of $|\alpha|$), it follows that  $\nu_\alpha^{\beta}$ and $\nu_\alpha^{\gamma}$  are siblings with parent~$\nu_\zeta^{\xi}$. 
	Then, there exist $i,j_1,j_2$ such that $\alpha=\zeta i$, $\beta=\xi j_1$ and $\gamma=\xi j_2$. However, according to Theorem~\ref{thm:incg}, the vertices in $A_{\zeta i}^{\xi j_1}$ and 
	$A_{\zeta i}^{\xi j_2}$ are at distance at least $5$, contradicting our assumption.
\end{proof}


\begin{lemma}
	\label{lem:rec}
	Let $\mathscr C$ be a monadically stable class. Assume that for each graph $G\in\mathscr C$  we have fixed a decomposition tree (with associated coloring $\coloring$ and maximum height $h$). 
	Let $\coloring(\mathscr C)=\{\coloring(v)\colon v\in V(G), G\in\mathscr C\}$,
	let $p$ be an integer, and let $\Gamma$ be a subset of $\coloring(\mathscr C)$ with size~$p$. 
	
	Then there exists a $\sigma$-expansion of~$\mathscr C$ (for some finite set $\sigma$ of unary predicates) such that, for every vertex $v$ with  $\coloring(v)\in\Gamma$ and every (strict) prefix $\alpha$ of $\coloring(v)$, the unique set $A_\alpha^{\beta}$ containing $v$ and the graph $G_\alpha^{\beta}$ are definable in the $\sigma$-expansion of $G$ with~$v$ as a parameter.
\end{lemma}
\begin{proof}
	Let $\Gamma^+$ be the set of all the prefixes of the words in $\Gamma$.
	We define, for each $\alpha\in\Gamma^+$, predicates $V_\alpha,F_\alpha$, and $A_\alpha$ as follows: $V_\alpha:=\bigcup_{\beta} V(G_\alpha^{\beta})$, $A_\alpha=\bigcup_{\beta} A_\alpha^{\beta}$,  $F_\alpha:=\bigcup_{\beta} F_\alpha^\beta$, and $Z_\alpha$ is the set of all the handles of the components  of the form $C_{\alpha}^{\beta}$.
	Let $\sigma$ be the set of all these predicates.
	As we introduced four predicates for each prefix $\alpha$ of a word in $\Gamma$, the number of predicates  in $\sigma$ is at most $4h|\Gamma|=4hp$.
	
	We prove the definability of the set 
	 $A_\alpha^{\beta}$ containing $v$ and of the graph $G_\alpha^{\beta}$
	 by induction on $|\alpha|$. If $|\alpha|=0$ (thus $\alpha=\beta=\lambda$), then 
	 $G_\lambda^\lambda=G$ and~$A_\lambda^\lambda=V(G)$ are obviously definable.
	
	Now assume $G_\alpha^{\beta}$ and $A_\alpha^{\beta}$ are definable in the $\sigma$-expansion of $G$ with parameter $v$, and let $i\in\mathbb N$ be such that $\alpha i$ is a prefix of $\coloring(v)$. 
	The set $A_{\alpha i}^{\beta}$ is definable as $A_{\alpha}^{\beta}\cap A_{\alpha i}$.

	Let $j$ be such that $v\in A_{\alpha i}^{\beta j}$.
	Then the handle $z_{\alpha i}^{\beta j}$ of $A_{\alpha i}^{\beta j}$ is at distance at most $2$ from all the vertices $A_{\alpha i}^{\beta j}$, including $v$. Moreover, if $\gamma\neq\beta j$ and $|\gamma|=|\beta|+1$, then $z_{\alpha i}^{\beta j}$ is at distance at least $3$ from the vertices in $A_{\alpha i}^\gamma$, for otherwise some vertex in  $A_{\alpha i}^\gamma$ would be at distance at most $4$ from some vertex in $A_{\alpha i}^{\beta j}$, contradiction Fact~\ref{fact:distA}. Thus, $z_{\alpha i}^{\beta j}$ 
	is the only vertex in~$Z_{\alpha i}$ at distance at most $2$ from $v$ in $G_\alpha^\beta$, and we have 
	$A_{\alpha i}^{\beta j}=A_{\alpha_i}\cap N_{G_\alpha^\beta}^2[z_{\alpha i}^{\beta j}]$. Hence, $A_{\alpha i}^{\beta j}$ is definable, as well as 
	$V(G_{\alpha_i}^{\beta j})=G_{\alpha}^\beta[N_{G_\alpha^\beta}[A_{\alpha i}^{\beta j}]]$.
	As $F_{\alpha i}^{\beta j}\subseteq V(G_{\alpha i}^{\beta j}$ and, according to Fact~\ref{fact:disj}, all the graphs $G_{\alpha i}^\gamma$ are vertex disjoint, we have
	$F_{\alpha i}^{\beta j}=F_{\alpha i}\cap V(G_{\alpha i}^{\beta j})$.
	Hence, $F_{\alpha i}^{\beta j}$ is also definable.
	
\end{proof}

\begin{theorem}
	\label{thm:qb}
	Every hereditary stable class of graphs $\mathscr C$ has quasi-bush representations by quasi-bushes in an almost nowhere dense class $\mathscr Q$. 
\end{theorem}
\begin{proof}
We construct a quasi-bush $Q$ representing $G$ from a decomposition tree by adding pointers from each vertex $\rho(u)$ (the leaf of the decomposition tree corresponding to $u$) to all the nodes $\nu_{\alpha}^{\beta}$ such that $u\in V(G_\alpha^{\beta})$.

Each arc $e=(\rho(u),\nu_\alpha^{\beta})$ is given a label, which is a binary vector $\mathbf b(e)$ of length $h$, computed as follows: the $\ell$th entry is $0$ if $\ell$ is greater than $|\alpha|$; 
otherwise, it is $1$ if $u$ belongs to $F_{\alpha[\ell]}^{\beta[\ell]}$, where $\alpha[\ell]$ and $\beta[\ell]$ are the prefixes of $\alpha$ and $\beta$ with length $\ell$. 

Similarly,  each vertex $v$ with $\rho(v)=\nu_\zeta^\xi$ is given a label; which is a binary vector $\mathbf b(v)$ whose $\ell$th entry is $1$ if $u$ belongs to $F_{\zeta[\ell]}^{\xi[\ell]}$, where $\zeta[\ell]$ and $\xi[\ell]$ are the prefixes of $\zeta$ and $\xi$ with length $\ell$. 

Now consider two distinct vertices $u,v$ of $G$ (See Fig.~\ref{fig:pointer}).
Let $e=(\rho(u),\nu_\alpha^\beta)$ be a pointer emanating from $\rho(u)$ such that $\nu_\alpha^\beta$ is an ancestor of $\rho(v)$ and $|\alpha|$ is maximal for this property (See Fig.~\ref{fig:pointer}). Such a pointer exists, as there is a pointer from $\rho(u)$ to the lowest common ancestor of $\rho(u)$ and $\rho(v)$ in the decomposition tree.

 By the maximality assumption,  $u$ does not belong to~$G_{\alpha i}^{\beta j}$.
As  $V(G_{\alpha i}^{\beta j})=N_{G_{\alpha}^{\beta}}[A_{\alpha i}^{\beta j}]$ and 
$v\in A_{\alpha i}^{\beta j}$ (as $\nu_{\alpha i}^{\beta j}$ is an ancestor of $\rho(v)$ in the decomposition tree), we deduce from $u\notin V(G_{\alpha i}^{\beta j})$ that $u$ is not adjacent to $v$ in $G_{\alpha}^{\beta}$. It follows, by applying all the flips going up to the root, that $u$ and $v$ are adjacent in $G$ if and only if $\langle \mathbf b(e),\mathbf b(v)\rangle=1$. This shows that~$Q$ is indeed a quasi-bush representation of $G$.

\begin{figure}[ht]
\begin{center}
	\includegraphics[width=\textwidth]{pointer}
	\caption{The quasi-bush $Q$. The lowest pointer $e$ from $\rho(u)$ to an ancestor of $\rho(v)$ is at worse the lowest common ancestor of $\rho(u)$ and $\rho(v)$ in the decomposition tree.}
	\label{fig:pointer}
\end{center}
\end{figure}

Now we prove that the class $\mathscr Q$ of the quasi-bushes we construct from graphs in $\mathscr C$ is almost nowhere dense.
According to  \cref{lem:hor_and}, it is sufficient to prove that $\mathscr Q$ has quasibounded-size nowhere dense decompositions.
In our construction of a quasi-bush $Q$ representing a graph $G\in\mathscr C$, we consider a decomposition tree $T$ of $G$ obtained by iteratively applying Theorem~\ref{thm:incg}. 
The number of words $\alpha\in\mathbb N^{<\omega}$ is easily seen to be in $\mathcal O_{\mathscr C,\epsilon}(n^{h\epsilon})$, where $n=|G|$, $\epsilon>0$ is the positive real considered in the applications of Theorem~\ref{thm:incg}, and $h\leq\Frank_6(G)\leq \Frank_6(\mathscr C)$ is the height of the decomposition tree.

Consider $G\in\mathscr C$, its decomposition tree $T$, and the derived quasi-bush representation $Q$ of $G$.
We extend the coloring $\coloring$ of the leaves of $T$
to all the vertices of $T$ (hence of $Q$) by defining $\coloring(\nu_{\alpha}^{\beta})=\alpha$. Let $W$ be the set of all the words $\alpha$ such that $T$ has a node with first index $\alpha$ (that is, of the form $\nu_\alpha^\beta$).
Now consider a subset $I$ of $W$ with size $p$. 
We define a set~$J$ of size at most $hp$ as follows:
for each maximal $\alpha\in I$, we put in $J$  all the words in $W$ of the form $\alpha 1\dots 1$ and we close $J$ by prefixes.
Let $\Gamma$ be the set of all the $\alpha$ in $J$ that are colors of vertices in $G$ (that is, of leaves of $T$).
According to Lemma~\ref{lem:rec} it is possible to construct the quasi-bush defined by the vertices with color in $\Gamma$ as a transduction $\mathsf T$ from $G$.
Note that for every node $\nu_{\alpha}^{\beta}$ of $Q$, either $\nu_{\alpha}^{\beta}$ is a leaf, or
$\nu_{\alpha 1}^{\beta 1}$ is a child of $\nu_{\alpha}^{\beta}$. It follows that the subgraph of~$Q$ induced by the vertices with color in $I$ is an induced subgraph of the quasi-bush obtained by $T$. 
Hence, there exists a class $\mathscr Q_p$ that can be obtained by transduction from $\mathscr C$ such that every  subgraph induced by $p$ colors of a quasi-bush constructed from a graph in $\mathscr C$  belongs to $\mathscr Q_p$. Thus, $\mathscr Q_p$ is monadically stable. By orienting the edges of a quasi-bush with $hp$ colors to the root, we get that every vertex has out-degree at most $hp$ (including one to its parent). Thus, $\mathscr Q_p$ is weakly sparse. 
Being both monadically stable and weakly sparse,
$\mathscr Q_p$ is nowhere dense \cite{msrw}.
Hence, $\mathscr Q$ has quasibounded-size nowhere dense decompositions, thus  (as mentioned above) $\mathscr Q$ is almost nowhere dense.
\end{proof}
\section{Proof of the main result and some consequences}
\label{sec:main}
We now prove Theorem~\ref{thm:stSD}, our main result.
Let us recall its statement.

\setcounter{tmpthm}{\value{theorem}}
\setcounter{theorem}{\value{thmmain}}
\begin{theorem}
	A hereditary class of graphs $\mathscr C$ is stable if and only if it has quasibounded-size bounded shrubdepth decompositions.	
\end{theorem}
\setcounter{theorem}{\value{tmpthm}}

Before proceeding, we specify how we encode a quasi-bush as a relational structure. 
The tree $T$ is encoded as a binary relation $\pi(x,y)$, such that 
$\pi(x,y)$ holds if $x$ is not the root and~$y$ is the parent of $x$, or $x$ is the root and $y=x$.
The leaves of $T$ are marked by a special predicate $L$, and their  labels are encoded by predicates $L_0,\dots,L_{h-1}$, where $L_i(x)$ means that the $i$th entry of $\mathbf b(x)$ is $1$.
The pointers and their labels are encoded by binary relations $P,P_0,\dots,P_{h-1}$. The existence of a pointer from $x$ to $y$ corresponds to $P(x,y)$, while $P_i(x,y)$ means that the $i$th entry of the label of this pointer is $1$. (Note that if the label has only null entries, then we have $P(x,y)\wedge\bigwedge_{0\leq i<h}\neg P_i(x,y)$.) This encoding will be used in the next proof.



\begin{proof}
	That a hereditary class with quasibounded-size bounded shrubdepth decompositions is stable follows from Theorem~\ref{thm:horizons}.
	
	Now consider a hereditary stable class of graphs $\mathscr C$ and an almost nowhere dense class $\mathscr Q$ of quasi-bush representations, whose existence is asserted by Theorem~\ref{thm:qb}.
\begin{claim}
	There is an existential interpretation $\mathsf I$ such that every graph $G\in\mathscr C$ is the $\mathsf I$-interpretation of its quasi-bush representation $Q$ in $\mathscr Q$.
\end{claim}
\begin{claimproof}
	Let $Q$ be the quasi-bush representation of $G\in\mathscr C$.
	The interpretation $\mathsf I$ constructs a graph, whose vertex set is the set of leaves of the tree $T$ of the quasi-bush $Q$ (that is the set of the vertices marked by the predicate $L$), and whose adjacency relation is defined as follows:

\begin{multline*}
E(x,y):=\exists  a_1,\dots,a_h\ \biggl((a_1=x)\wedge\bigwedge_{1\leq i<h}\pi(a_i,a_{i+1})\ \wedge \\
\bigvee_{1<i\leq h} \Bigl(P(x,a_i)\wedge\bigwedge_{1<j<i}\neg P(x,a_j)\wedge
{\rm Odd}(P_0(x,a_i)\wedge L_0(y),\dots,P_{h-1}(x,a_i)\wedge L_{h-1}(y))\Bigr)\biggr),
\end{multline*}
where ${\rm Odd}(x_1,\dots,x_h)$ is the Boolean function, whose value is true if an odd number of its arguments are true.
\end{claimproof}

Fix a parameter $p\in\mathbb N$ and $\epsilon>0$.
As  $\mathscr Q$ is almost nowhere dense,  it has quasibounded-size bounded treedepth decompositions. 
Hence, there exists a function $N:\mathbb N\rightarrow \mathbb N$ such that $N(n)\in \mathcal O(n^\epsilon)$ for every $\epsilon>0$ and, for every $Q\in\mathscr Q$, there exists  a coloring 
 Let $c:V(Q)\rightarrow [N(|Q|)]$, such that every $p$ color classes induce a substructure in a class $\mathscr Q_p$ with bounded treedepth, which we can assume to be hereditary.
 
Let $Q\in\mathscr Q$, let $c$ be the associated coloring, and let $G=\mathsf I(Q)$. For clarity, we denote by~$\rho$ the inclusion map from $V(G)$ to $V(Q)$. We define a coloring $\gamma$ on $V(G)$ by
\[
\gamma(v)=(c(\rho(v_1)),\dots,c(\rho(v_h)),
\]
where $v_1=v$ and $Q\models \pi(v_i,v_{i+1})$ (that is, $v_{i+1}$ is the parent of $v_i$, considering that the root is its own parent).
Let $\Gamma$ be the set of all the colors used to color $V(G)$ by $\gamma$. Note that $|\Gamma|\leq N(|Q|)^h\in \mathcal O(|Q|^\epsilon)$ for every $\epsilon>0$. Moreover, as $|Q|\leq h |G|$, we have $|\Gamma|\in \mathcal O(|G|^\epsilon)$ for every $\epsilon>0$ as well.

Let $I\subseteq \Gamma$ be a set of $p$ colors and let $A$ be the subset of all the vertices of $G$ with their $\gamma$-color in $I$.

Let $\rho(A)=\{\rho(v)\colon v\in A\}$ and let $B$ be the union of $A$ with the set of all the ancestors in~$T$ of vertices in $\rho(A)$. 
By construction, the $c$-colors of the vertices in $B$ belong to a set~$J$ of size at most $h|I|=hp$. Moreover, it is clear that $\mathsf I(Q[B])=\mathsf I(Q)[A]=G[A]$. Hence, $G[A]\in \mathsf I(\mathscr Q_p)$, which has bounded shrubdepth (being a class of interpretations of binary structures with bounded treedepth). 

Consequently, the class $\mathscr C$ has quasibounded-size bounded shrubdepth decompositions~\cite{dreier2022treelike}.
\end{proof}

\subsection{Erd\H os-Hajnal property}

A hereditary class of graphs $\mathscr C$ has the \emph{Erd\H os-Hajnal property} if there exists a constant $c > $0 such that every
graph $G\in\mathscr C$  has a clique or independent set of size $\Omega_{\mathscr C}(|G|^c)$. 
The celebrated Erd\H os-Hajnal conjecture asserts that for every fixed graph $H$,  the class of all graphs with no induced subgraph isomorphic to $H$ has the Erd\H os-Hajnal property.
It has been proved recently \cite{nguyen2023induced} that every 
hereditary class of graphs with bounded VC-dimension (hence, every hereditary dependent class) has the Erd\H os-Hajnal property.
Malliaris and Shelah  implicitly proved~\cite{Malliaris2014}
that hereditary classes of graphs with stable edge relation have the Erd\H os-Hajnal property by developing regularity lemmas for these graphs.  However, the derived exponent depends on the size of the largest half-graph present in the graphs of the class.
We prove that under the stronger assumption of stability, the exponent can be taken arbitrarily close to $1/2$. 
	
Following \cite{Gimbel20103437}, denote by $c(G)$ the \emph{c-chromatic number} of a graph $G$, that is, the minimum size of a partition of the vertex set of $G$ such that each part induces a cograph. 
As every cograph $G$ has a clique or independent set with size $|G|^{1/2}$, we immediately infer that for every graph $G$ we have
\[
\max(\alpha(G),\omega(G))\geq c(G)^{-1/2}\,|G|^{1/2}.
\]

\begin{remark}
	\label{rem:lcw}
For instance, as the c-chromatic number of a graph is bounded by a function of its linear clique-width \cite{msrw}, there is a function $f:\mathbb N\rightarrow (0,1]$ such that every graph $G$ with linear clique-width $t$ has a clique or an independent set of size $f(t)\,|G|^{1/2}$.
\end{remark}

As every class of graphs with bounded shrubdepth has bounded c-chromatic number (see \cite{msrw}) we get that every hereditary class $\mathscr C$ with quasibounded-size bounded shrubdepth decompositions (at least for parameter $1$) has the Erd\H os-Hajnal property with exponent $1/2-\epsilon$, for every $\epsilon>0$.
In particular, we have.

\begin{corollary}
	\label{cor:EH}
	Every graph $G$ in a hereditary stable class $\mathscr C$ has a clique or an independent set of size $\Omega_{\mathscr C,\epsilon}(|G|^{1/2-\epsilon})$ for every $\epsilon>0$.
\end{corollary}

Note that the exponent can be set to $1/2$ instead of $1/2-\epsilon$ for classes with structurally bounded expansion, as they have bounded-size bounded-shrubdepth decompositions, and this cannot be improved 
(even for classes with shrubdepth at most $2$) as witnessed by the class of all disjoint union of cliques, as it contains graphs on $n$ vertices ($\sqrt{n}$ copies of $K_{\sqrt{n}}$) with no independent set or clique of size greater than $\sqrt{n}$.
 However, for general hereditary stable classes, the bound of Corollary~\ref{cor:EH} is tight is the following sense.

\begin{proposition}
	Let $N:\mathbb N\rightarrow\mathbb N$.  There exists a hereditary stable class~$\mathscr C$ such that for every sufficiently large integer $c$, there exists in $\mathscr C$ a graph $G$ of order at least~$N(c)$ with $\max(\alpha(G),\omega(G))<|G|^{1/2-1/c}$. 
	
	Consequently, the bound $\Omega_{\mathscr C,\epsilon}(|G|^{1/2-\epsilon})$ for every $\epsilon>0$ given in Corollary~\ref{cor:EH} cannot be improved to $\Omega_{\mathscr C}(|G|^{1/2})$.
\end{proposition}	
\begin{proof}
	Let $c\in\mathbb N$ be sufficiently large.
	Let $p$ be a prime such that
	$p\equiv 1\bmod 4$ and 
	$p>2N(c)^{1/c}$.
	Let $q$ be a prime such that 
	$q\equiv 1\bmod 4p$ and 
	$q$ is slightly smaller than $p^{c/6}$ (existence follows from the estimation of  the density of primes in an arithmetic progression proved by de~la~Vall\'ee Poussin). Then $q\equiv 1\bmod 4$ and $q$ is a quadratic residue modulo~$p$. (Hence $p$ is a quadratic residue modulo $p$ according to the quadratic residue theorem.) The construction of 
	Lubotzky, Phillips and Sarnak~\cite{Lubotzky1988a} provides a graph $X^{p,q}$ with 
	the following properties:
$X^{p,q}$ is $p+1$-regular, has order $n=q(q^2-1)/2>N(c)^{1/2}$,
 girth	at least $2 \log_p q\approx c/3$, and independence number $\alpha(X^{p,q})=2 \sqrt{p}/(p+1) n \approx n^{1/2-2/c}$.
We define 
$G_c=\overline{X^{p,q}}\bullet X^{p,q}$, where $\bullet$ denotes the lexicographic product. We further define $H_c$ as the graph obtained from $X^{p,q}$ by joining each vertex $v$ of $X^{p,q}$ to all the vertices of a private copy of $X^{p,q}$. Note that 
$|G_c|=n^2+n$, $|H_c|=n^2> N(c)$ and
that $\max(\alpha(G_c),\omega(G_c))=2\alpha(X^{p,q})\approx |G_c|^{1/2-1/c}$.

Let $\mathscr D=\{H_c: c\geq 9\}$ and  let $k$ be an integer. Assume that the $k$-th subdivision of $K_n$ is a subgraph of some graph $H_c\in\mathscr D$. Being $2$-connected, it is a subgraph of a block of some~$H_c$, thus the $k$-th subdivision of $K_{n-1}$ is a subgraph of some graph $X^{p,q}$ with girth at least~$c/3$. It follows that $c\leq 9k$ hence
$n\leq \max_{c\leq 9k}|H_c|$. 
We deduce that $\mathscr D$ is nowhere dense.  Let $\mathscr C$ be the hereditary closure of $\{G_c: c\geq 9\}$. As $\mathscr C$ can be  obtained as a transduction of $\mathscr D$, it is stable, and it satisfies the conditions of the Proposition by construction.
\end{proof}

\begin{figure}[h!t]
	\begin{center}
		\includegraphics[width=\textwidth]{constr}
		\caption{Construction of the graphs $H_c$ and $G_c$ (in the case $X^{p,q}$ is $C_5$).}
	\end{center}
\end{figure}

\subsection{Logarithmic density of induced subgraphs}
The arguments in the proofs given in \cite{Taxi_hom} 
can be followed \emph{mutatis mutandis} (by considering induced subgraphs of graphs with bounded shrubdepth instead of subgraphs of graphs with bounded treedepth), and we get
\begin{corollary}
	\label{cor:count}
	For every hereditary stable class $\mathscr C$ we have
\[
	\limsup_{G\in\mathscr C}\frac{\log (\# F\subseteq_i G)}{\log |G|}\in\mathbb N\cup\{-\infty\}.
\]
\end{corollary}

Precisely, following  \cite{Taxi_hom}, one proves that  if $\limsup_{G\in\mathscr C}\log (\# F\subseteq_i G)/\log |G|>k$, then 
there exists, for every integer $n\in\mathbb N$ a graph $G_n\in\mathscr C$ and a partition 
$\{V_v\colon v\in V(F)\}$ of~$V(G_n)$ indexed by~$V(F)$ such that 
\begin{itemize}
	\item for each $v\in F|$, either $|V_v|=1$ or $|V_v|=n$;
	\item there are at least $k+1$ vertices $v\in V(F)$ with  $|V_v|=n$;
	\item each $V_i$ is either a clique or an independent set;
	\item every function $f:V(F)\rightarrow V(G_n)$ with $f(v)\in V_v$ defines a isomorphism of $F$ and its image.
\end{itemize}
Corollary~\ref{cor:count} then follows.

\section{Discussion}
We have proved that several hereditary graph properties are decomposition horizons. We have chosen to give direct proofs for some of them where we instead could have used  \cref{cor:int} and later results.

\pagebreak
For instance:
\begin{itemize}
\item ``nowhere dense'' is a decomposition horizon, as it is the intersection of the decomposition horizons ``stable'' and ``weakly sparse'';
\item ``bounded maximum degree after deletion of a bounded number of vertices'' is a decomposition horizon,  as it is the intersection of ``transduction of a class with bounded degree'' and weakly sparse.
\end{itemize}

Also, we would like to stress that  general properties of decomposition horizons and ``dependent'' and ``stable'' are decomposition horizons hold in the more general setting of relational structures.

Our examples include an infinite countable chain of decomposition horizons (\cref{cor:nearly bounded degree}), witnessing some richness of the inclusion order of decomposition horizons. Note that this order is a distributive lattice (by \Cref{cor:int,rem:union}).
\medskip

We now consider few problems naturally arising from our study.
\medskip

In our study of decomposition horizons, we have introduced the operators $\Pi\mapsto\Pi^+$ and~$\Pi\mapsto\Pi^\ast$. While the former is clearly idempotent (cf.\  Corollary~\ref{cor:idem}), it is not clear whether the latter is.
\begin{problem}
	Is there a hereditary (dependent) class property $\Pi$ such that $\Pi^\ast\neq(\Pi^\ast)^\ast$?
\end{problem}

Also, we have seen that the operators $\Pi\mapsto\Pi^+$ and $\Pi\mapsto\Pi^\ast$ preserve finite intersections (Corollary~\ref{cor:int}).
This suggests the next problem.

\begin{problem}
	\label{pb:min}
	Is it true that for every hereditary class property $\Pi$ there is an inclusion-minimum hereditary class  property $\Lambda$ with $\Lambda^+=\Pi^+$?
	The same problem can be asked for
	$\Lambda^\ast=\Pi^\ast$.
\end{problem}

We will say that a hereditary class property $\Pi$ is \emph{minimal} if there is no smaller hereditary class property $\Lambda$  with $\Lambda^+=\Pi^+$ or $\Lambda^\ast=\Pi^\ast$. It is easily seen that for all integers $n,k,h,c$ there exists a $c$-colored forest with height $h$
with $(nkc)^h$ vertices such that in every partition of the vertex set into $k$ parts and every $c$-colored forest $F$ with height at most $h$, 
some copy of $F$ is included within $c$ parts. From this it follows that the hereditary class properties ``bounded treedepth'' and ``bounded shrubdepth'' are minimal.
Hence, for every hereditary class property $\Pi$, we have the following equivalences.
\begin{align*}
&\text{``bounded treedepth''}\subseteq \Pi\subseteq \text{``bounded expansion''}&\iff&\Pi^+=\text{``bounded expansion''}\\
&\text{``bounded treedepth''}\subseteq \Pi\subseteq \text{``nowhere dense''}&\iff&\Pi^\ast=\text{``nowhere dense''}\\
&\text{``bounded shrubdepth''}\subseteq \Pi\subseteq \text{\rlap{``structurally bounded expansion''}}&&\\
&\mathrlap{\qquad\qquad\qquad\qquad\qquad\qquad\qquad\qquad\qquad\iff\Pi^+=\text{``structurally bounded expansion''}}&&\\
&\text{``bounded shrubdepth''}\subseteq \Pi\subseteq \text{``stable''}&\iff&\Pi^\ast=\text{``stable''}
\end{align*}
\medskip

While stable hereditary classes of graphs are exactly those hereditary classes with quasibounded-size bounded shrub-depth decompositions, dependent hereditary classes seem to be more elusive.
It was proved in  \cite{tww_ordered} that 
for hereditary classes of ordered graphs, being dependent is equivalent to having bounded twin-width.
On the other hand,  classes with quasibounded-size bounded twin-width decompositions are dependent (as classes with bounded twin-width are dependent) and include  all stable hereditary classes.
This suggests the following problem.

\begin{problem}
Is the class property ``dependent'' the decomposition horizon of the class property ``bounded twin-width''?
More generally,
	what is the smallest class property $\Pi$ such that every hereditary dependent class has a quasibounded $\Pi$-decomposition (assuming such a minimal class exists, cf.\  Problem~\ref{pb:min})?
\end{problem}

We remark that ``bounded linear cliquewidth'' is a too strong class property to have ``dependent'' as its decomposition horizon, as even classes with bounded cliquewidth do not have, in general, quasibounded-size bounded linear cliquewidth decompositions (even with parameter~1). 
Indeed, every hereditary class $\mathscr C$ with quasibounded-size bounded linear cliquewidth decompositions has the Erd\H os-Hajnal property with exponent $1/2$ (see Remark~\ref{rem:lcw}).
On the other hand,  the hereditary closure of the lexicographic powers of $C_5$ has bounded cliquewidth but
the graph $G=C_5^{\bullet n}$ (which belongs to  $\mathscr C$) has no homogeneous set of size greater than $|G|^{\log 2/\log 5}<|G|^{0.41}$.
(Note that a similar argument shows that there is no constant $\beta>0$ such that for every class $\mathscr C$   with bounded cliquewidth every graph $G$ in $\mathscr C$ has an independent set or a clique of size $\Omega_{\mathscr C}(|G|^\beta)$.)


\section*{Acknowledgments}
The authors wish to express their sincere gratitude to the referees for their careful reading and constructive suggestions.

\bibliography{ref}
\end{document}